\documentclass[ALICE,manyauthors]{cernphprep}
\usepackage[comma,square,numbers,sort&compress]{natbib}
\usepackage{hyperref}
\usepackage{lineno}
\usepackage{xspace}
\usepackage{color}
\usepackage{float}
\usepackage[T1]{fontenc}
\usepackage{orcidlink}
\begin{document}
%

\newcommand{\pp}           {pp\xspace}
\newcommand{\ppbar}        {\mbox{$\mathrm {p\overline{p}}$}\xspace}
\newcommand{\XeXe}         {\mbox{Xe--Xe}\xspace}
\newcommand{\PbPb}         {\mbox{Pb--Pb}\xspace}
\newcommand{\pA}           {\mbox{pA}\xspace}
\newcommand{\pPb}          {\mbox{p--Pb}\xspace}
\newcommand{\AuAu}         {\mbox{Au--Au}\xspace}
\newcommand{\dAu}          {\mbox{d--Au}\xspace}
\newcommand{\CuCu}         {\mbox{Cu--Cu}\xspace}

\newcommand{\s}            {\ensuremath{\sqrt{s}}\xspace}
\newcommand{\snn}          {\ensuremath{\sqrt{s_{\mathrm{NN}}}}\xspace}
\newcommand{\pt}           {\ensuremath{p_{\rm T}}\xspace}
\newcommand{\meanpt}       {$\langle p_{\mathrm{T}}\rangle$\xspace}
\newcommand{\ycms}         {\ensuremath{y_{\rm CMS}}\xspace}
\newcommand{\ylab}         {\ensuremath{y_{\rm lab}}\xspace}
\newcommand{\etarange}[1]  {\mbox{$\left | \eta \right |~<~#1$}}
\newcommand{\yrange}[1]    {\mbox{$\left | y \right |~<$~0.5}}
\newcommand{\dndy}         {\ensuremath{\mathrm{d}N_\mathrm{ch}/\mathrm{d}y}\xspace}
\newcommand{\dndeta}       {\ensuremath{\mathrm{d}N_\mathrm{ch}/\mathrm{d}\eta}\xspace}
\newcommand{\avdndeta}     {\ensuremath{\langle\dndeta\rangle}\xspace}
\newcommand{\dNdy}           {\ensuremath{\mathrm{d}N_\mathrm{ch}/\mathrm{d}y}\xspace}
\newcommand{\dNdyy}         {\ensuremath{\mathrm{d}N/\mathrm{d}y}\xspace}
\newcommand{\Npart}        {\ensuremath{N_\mathrm{part}}\xspace}
\newcommand{\Ncoll}        {\ensuremath{N_\mathrm{coll}}\xspace}
\newcommand{\dEdx}         {\ensuremath{\textrm{d}E/\textrm{d}x}\xspace}
\newcommand{\RpPb}         {\ensuremath{R_{\rm pPb}}\xspace}
\newcommand{\RAA}         {\ensuremath{R_{\rm AA}}\xspace}

\newcommand{\nineH}        {$\sqrt{s}~=~0.9$~Te\kern-.1emV\xspace}
\newcommand{\seven}        {$\sqrt{s}~=~7$~Te\kern-.1emV\xspace}
\newcommand{\twoH}         {$\sqrt{s}~=~0.2$~Te\kern-.1emV\xspace}
\newcommand{\twosevensix}  {$\sqrt{s}~=~2.76$~Te\kern-.1emV\xspace}
\newcommand{\five}         {$\sqrt{s}~=~5.02$~Te\kern-.1emV\xspace}
\newcommand{\twosevensixnn}{$\sqrt{s_{\mathrm{NN}}}~=~2.76$~Te\kern-.1emV\xspace}
\newcommand{\fivenn}       {$\sqrt{s_{\mathrm{NN}}}~=~5.02$~Te\kern-.1emV\xspace}
\newcommand{\LT}           {L{\'e}vy-Tsallis\xspace}
\newcommand{\GeVc}         {Ge\kern-.1emV/$c$\xspace}
\newcommand{\MeVc}         {Me\kern-.1emV/$c$\xspace}
\newcommand{\TeV}          {Te\kern-.1emV\xspace}
\newcommand{\GeV}          {Ge\kern-.1emV\xspace}
\newcommand{\MeV}          {Me\kern-.1emV\xspace}
\newcommand{\GeVmass}      {Ge\kern-.2emV/$c^2$\xspace}
\newcommand{\MeVmass}      {Me\kern-.2emV/$c^2$\xspace}
\newcommand{\lumi}         {\ensuremath{\mathcal{L}}\xspace}

\newcommand{\ITS}          {\rm{ITS}\xspace}
\newcommand{\TOF}          {\rm{TOF}\xspace}
\newcommand{\ZDC}          {\rm{ZDC}\xspace}
\newcommand{\ZDCs}         {\rm{ZDCs}\xspace}
\newcommand{\ZNA}          {\rm{ZNA}\xspace}
\newcommand{\ZNC}          {\rm{ZNC}\xspace}
\newcommand{\SPD}          {\rm{SPD}\xspace}
\newcommand{\SDD}          {\rm{SDD}\xspace}
\newcommand{\SSD}          {\rm{SSD}\xspace}
\newcommand{\TPC}          {\rm{TPC}\xspace}
\newcommand{\TRD}          {\rm{TRD}\xspace}
\newcommand{\VZERO}        {\rm{V0}\xspace}
\newcommand{\VZEROA}       {\rm{V0A}\xspace}
\newcommand{\VZEROC}       {\rm{V0C}\xspace}
\newcommand{\Vdecay} 	   {\ensuremath{V^{0}}\xspace}

\newcommand{\ee}           {\ensuremath{e^{+}e^{-}}} 
\newcommand{\pip}          {\ensuremath{\pi^{+}}\xspace}
\newcommand{\pim}          {\ensuremath{\pi^{-}}\xspace}
\newcommand{\kap}          {\ensuremath{\rm{K}^{+}}\xspace}
\newcommand{\kam}          {\ensuremath{\rm{K}^{-}}\xspace}
\newcommand{\pbar}         {\ensuremath{\rm\overline{p}}\xspace}
\newcommand{\kzero}        {\ensuremath{{\rm K}^{0}_{\rm{S}}}\xspace}
\newcommand{\lmb}          {\ensuremath{\Lambda}\xspace}
\newcommand{\almb}         {\ensuremath{\overline{\Lambda}}\xspace}
\newcommand{\Om}           {\ensuremath{\Omega^-}\xspace}
\newcommand{\Mo}           {\ensuremath{\overline{\Omega}^+}\xspace}
\newcommand{\X}            {\ensuremath{\Xi^-}\xspace}
\newcommand{\Ix}           {\ensuremath{\overline{\Xi}^+}\xspace}
\newcommand{\Xis}          {\ensuremath{\Xi^{\pm}}\xspace}
\newcommand{\Oms}          {\ensuremath{\Omega^{\pm}}\xspace}
\newcommand{\degree}       {\ensuremath{^{\rm o}}\xspace}
\newcommand{\kstar}        {\ensuremath{\rm {K}^{\rm{* 0}}}\xspace}
\newcommand{\phim}        {\ensuremath{\phi}\xspace}
\newcommand{\pik}          {\ensuremath{\pi\rm{K}}\xspace}
\newcommand{\kk}          {\ensuremath{\rm{K}\rm{K}}\xspace}
\newcommand{\kskm}{$\mathrm{K^{*0}/K^{-}}$}
\newcommand{\phikm}{$\mathrm{\phi/K^{-}}$}
\newcommand{\phixi}{$\mathrm{\phi/\Xi}$}
\newcommand{\phiom}{$\mathrm{\phi/\Omega}$}
\newcommand{\xiphi}{$\mathrm{\Xi/\phi}$}
\newcommand{\omphi}{$\mathrm{\Omega/\phi}$}
\newcommand{\kstf} {K$^{*}(892)^{0}~$}
\newcommand{\phf} {$\mathrm{\phi(1020)}~$}
\newcommand{\dd}{\ensuremath{\mathrm{d}}}
\newcommand{\mT}{\ensuremath{m_{\mathrm{T}}}\xspace}
\newcommand{\krr}{\ensuremath{\kern-0.09em}}

\begin{titlepage}
\PHyear{2022}       
\PHnumber{067}      
\PHdate{24 March}  

\title{Multiplicity and rapidity dependence of $\mathrm{K}^{*}(\mathrm{\textbf{892}})^{0}$ and $\mathrm{\phi(\textbf{1020})}$ ~~~~~~~~~~~~~ ~~~~~~~~~~~~~ production in p--Pb collisions at $\snn =$ 5.02 \TeV }
\ShortTitle{$\mathrm{K}^{*}(\mathrm{{892}})^{0}$ and $\mathrm{\phi({1020})}$ in p--Pb at $\snn =$ 5.02 \TeV }   
\Collaboration{ALICE Collaboration\thanks{See Appendix~\ref{app:collab} for the list of collaboration members}}
\ShortAuthor{ALICE Collaboration} 
\begin{abstract}
The transverse-momentum ($\pt$) spectra of \kstf and \phf measured with the ALICE detector up to  $\pt$ = 16 \GeV/$c$ in the rapidity range $-$1.2 $<$ $y$ $<$ 0.3, in p--Pb collisions at the center-of-mass energy per nucleon--nucleon collision \mbox{$\snn = 5.02$ \TeV} are presented as a function of charged particle multiplicity and rapidity.
The measured $\pt$ distributions show a dependence on both multiplicity and rapidity at low $\pt$ whereas no significant dependence is observed at high $\pt$. A rapidity dependence is observed in the $\pt$-integrated yield (d$N$/d$y$), whereas the mean transverse momentum $\left(\langle\pt\rangle\right)$ shows a flat behavior as a function of rapidity.
The rapidity asymmetry ($Y_\mathrm{asym}$) at \mbox{low $\pt$ ( $<$ 5 \GeV/$c$)} is more significant for higher multiplicity classes.
At high $\pt$, no significant rapidity asymmetry is observed in any of the multiplicity classes. Both \kstf and \phf show similar $Y_\mathrm{asym}$. The nuclear modification factor ($Q_\mathrm{CP}$) as a function of $\pt$ shows a Cronin-like enhancement at intermediate $\pt$, which is more prominent at higher rapidities (Pb-going direction) and in higher multiplicity classes. At high \pt ($>$ 5 \GeVc), the $Q_\mathrm{CP}$ values are greater than unity and no significant rapidity dependence is observed.	    
\end{abstract}
\end{titlepage}
\setcounter{page}{2} 

\section{Introduction}
The primary goals of high-energy heavy-ion (A--A) collisions are to create a system of deconfined quarks and gluons known as quark--gluon plasma (QGP) and to study its properties~\cite{Gyulassy:2004zy,Adams:2005dq,Schukraft:2011na,Braun-Munzinger:2015hba}.
Asymmetric collision systems like proton-nucleus (p--A) and deuteron-nucleus (d--A) can be considered as control experiments where the formation of an extended QGP phase is not expected. These collision systems are used as
\mbox{baseline} measurements to study the possible effects of cold nuclear matter and disentangle the same from hot dense matter effects produced in heavy-ion 
collisions~\cite{PHENIX:2003qdw,PHOBOS:2003uzz,BRAHMS:2004xry,ZEUS:1997etp,ALICE:2012mj,Acharya:2018qsh, Adams:2004ep,Anticic:2011zr,Acharya:2019yoi,ALICE:2021ptz}. 
In  addition, p--A collisions at Large Hadron Collider (LHC) energies enable probing the parton distribution functions in nuclei at very small values of the Bjorken $x$ variable, where gluon saturation effects may occur~\cite{Albacete:2016veq,CMS:2016zzh,CMS:2019isl}. Recent measurements in high-multiplicity pp, $\pPb$, p--Au, d--Au, and $^{3}$He--Au collisions at different energies have shown features such as anisotropies in particle emission azimuthal angles, strangeness enhancement, and  long--range structures in two-particle angular correlations on the near and away side,
 which previously have been observed in nucleus-nucleus collisions~\cite{STAR:2006kxj,PHENIX:2018hho,PHENIX:2014fnc,PHENIX:2016cfs,Kovchegov:2004jm,ALICE:2016fzo,ALICE:2012eyl,PHENIX:2018lia,Adam:2016bpr,ALICE:2021uyz,ALICE:2021ucq,ALICE:2015cql}. The origin of these phenomena in small systems is not yet fully understood. A systematic study of multiplicity and rapidity dependence of hadron production allows us to investigate the mechanism of particle production and shed light on the physics processes that contribute to the particle production~\cite{Albacete:2016veq}. Similar studies have been reported by the experiments at the LHC~\cite{CMS:2016zzh,CMS:2019isl,ALICE:2012mj,Acharya:2018qsh} and Relativistic Heavy Ion Collider (RHIC)~\cite{PHENIX:2018hho,PHENIX:2014fnc,PHOBOS:2003uzz,PHENIX:2016cfs,BRAHMS:2004xry}. 
The mechanism of hadron production may be influenced by different effects such as nuclear modification of the parton distribution functions (nuclear shadowing) and possible parton saturation, multiple scattering, and radial flow ~\cite{Kang:2012kc,ALICE:2013wgn,Bozek:2015swa,CMS:2016zzh}. These effects are expected to depend on the rapidity of the produced particles. In \pPb collisions, one can expect that the production mechanism may be sensitive to different effects at forward (p-going) and backward (Pb-going) rapidities~\cite{CMS:2016zzh,CMS:2019isl,ALICE:2012mj,Acharya:2018qsh,Bozek:2013sda,Bozek:2015swa}. 
The partons of the incident proton are expected to undergo multiple scattering while traversing the Pb-nucleus. 
It is thus interesting to study the ratio of particle yields between Pb- and p-going directions, represented by the rapidity asymmetry ($Y_\mathrm{asym}$) defined as:
   \begin{equation}
  Y_\mathrm{asym}(p_\mathrm{T}) =\frac{\left.\frac{\mathrm{d}^{2}N}{\mathrm{d}p_\mathrm{T}\mathrm{d}y}\right\vert_{-0.3 < y < 0}}{\left.\frac{\mathrm{d}^{2}N}{\mathrm{d}p_\mathrm{T}\mathrm{d}y}\right\vert_{0 < y < 0.3}}
 \label{eqn:asym}
 \end{equation}
 where 
  $\rm{d}^{2}\it{N}/ \rm{d}\it{p}_{\rm{T}}\rm{d}\it{y}|_{-\mathrm{0.3} < y < \mathrm{0}}$ is the particle yield in the rapidity ($y$) interval --0.3 $< y < $ 0, considered as the Pb-going direction, and
  $\rm{d}^{2}\it{N}/\rm{d}\it{p}_{\rm{T}}\rm{d}\it{y}|_{\mathrm{0} < y < \mathrm{0.3}}$ 
    is the particle yield in the rapidity interval \mbox{0 $< y < $ 0.3}, corresponding to the p-going direction. 
 From the experimental point of view, the $Y_\mathrm{asym}$ is a powerful observable because systematic uncertainties cancel out in the ratio and hence it can better discriminate rapidity-dependent effects among models~\cite{STAR:2006kxj,CMS:2019isl,CMS:2016zzh}. 
 Gluon saturation effects at low Bjorken $x$ values~\cite{BRAHMS:2004xry,STAR:2006kxj} may affect the transverse momentum distribution of hadron production at large rapidities in the p-going direction in \pPb collisions at LHC energies. The gluon saturation effects depend on the colliding nuclei and rapidity as A$^{1/3}e^{\lambda y}$, where $A$ represents the mass number~\cite{STAR:2006kxj}, and $\lambda$ is a \mbox{parameter} whose value lies between\mbox{ 0.2 and 0.3,} and is obtained from fits to the HERA measurements~\cite{ZEUS:1997etp}. The effect of rapidity dependence on particle production is tested by measuring the ratios of integrated yield (d$N/$d$y$) and mean transverse momentum  $\left(\langle\pt\rangle\right)$ at given $y$ to the values at $y=0$, i.e.\ denoted as (d$N/$d$y$)/(d$N/$d$y)_{y=0}$ and $\langle\pt\rangle$/$\langle\pt\rangle_{y=0}$. 
 It is also important to study the 
 variation with rapidity of the nuclear modification factor between central and non-central collisions.  
 This factor $\left(Q_{\mathrm{CP}} (p_\mathrm{T})\right)$ is defined as
   \begin{equation}
      \it{Q}_{\mathrm{CP}} (p_\mathrm{T})=  \left.\frac{\frac{\mathrm{d}^{2}N}{\mathrm{d}p_\mathrm{T}\mathrm{d}y}}{\langle N_\mathrm{coll}\rangle}\right\vert_{\textrm{HM}}  \Bigg / \left. \frac{\frac{\mathrm{d}^{2}N}{\mathrm{d}p_\mathrm{T}\mathrm{d}y}}{\langle N_\mathrm{coll} \rangle}\right\vert_{\textrm{LM}},
 \label{eqn:qcp}
 \end{equation}
 
 where $\langle N_\mathrm{coll} \rangle$ is the average number of nucleon--nucleon collisions in low-multiplicity (\textrm{LM}) and high-multiplicity (\textrm{HM}) events, respectively.
The multiplicity dependence of \kstar and \phim meson production at midrapidity was studied in pp, \pPb, and \PbPb collisions at LHC energies and reported in Refs.~\cite{Adam:2016bpr,Acharya:2019bli,ALICE:2021uyz,ALICE:2021ptz}. The lifetime of the \kstar meson is about 4 fm/$c$, which is comparable to the lifetime of the hadronic phase. In contrast, the lifetime of the \phim meson is 10 times higher. The \kstar and \phim mesons are thus useful probes of the late-stage evolution of high-energy hadronic collisions. As they have similar mass but differ in their strangeness content by one unit, they are suitable candidates to understand the modification of particle production due to rescattering effects and the role of the strangeness content, as discussed in Refs.~\cite{ALICE:2015cql,Adam:2016bpr,Acharya:2019bli,ALICE:2021ucq}.
 
This article reports the first measurements of the rapidity dependence of \kstf and \phf mesons production in $\pPb$ collisions at center-of-mass energy per nucleon--nucleon collisions at \mbox{$\snn$ $=$ 5.02} \TeV by the ALICE experiment at the LHC. The large size of the data sample and the excellent particle identification (PID) provide opportunities to extend these measurements in a wider rapidity interval and multiplicity classes compared to earlier measurements~\cite{ALICE:2016fzo,ALICE:2012eyl,Adam:2016bpr,ALICE:2021uyz,ALICE:2021ucq,ALICE:2015cql}. This enables the investigation of the nuclear effects on the particle production in \pPb collisions. The \pt spectra,  $Y_\mathrm{asym}(p_\mathrm{T})$ and $\it{Q}_{\mathrm{CP}} (p_\mathrm{T})$ are studied  in the rapidity range -1.2 $< y <$ 0.3 and three multiplicity classes along with a measurement on the multiplicity-integrated sample. Similar measurements for charged particles and strange hadrons at RHIC and the LHC energies were reported in Refs.~\cite{BRAHMS:2004xry,STAR:2006kxj,CMS:2016zzh}.

The measurements presented here are compared with various model predictions such as EPOS-LHC~\cite{Pierog:2013ria}, EPOS3 with and without UrQMD~\cite{Werner:2013yia,Pierog:2009zt,Werner:2013tya,Knospe:2021jgt}, DPMJET~\cite{Roesler:2000he}, HIJING~\cite{Gyulassy:1994ew} and PYTHIA8/Angantyr~\cite{Bierlich:2018xfw}. EPOS-LHC is an event generator for minimum-bias hadronic interactions that incorporates a parameterization of flow based on LHC data~\cite{Pierog:2013ria}. It is an event generator based on multiple partonic scatterings described using Gribov's Reggeon field theory formalism, supplemented with collective hadronization and the core-corona mechanism from pp to A--A collisions~\cite{Pierog:2013ria}. EPOS3 is an event generator based on 3$+$1D viscous hydrodynamical evolution in the Gribov-Regge multiple scattering framework, which is used to understand hadronic resonances and their interactions in the partonic and hadronic medium. The UrQMD model introduces the description of rescattering and regeneration effects in the hadronic phase~\cite{Werner:2013yia,Werner:2013tya,Knospe:2021jgt}.
DPMJET is a QCD-inspired dual parton model based on the formalism of the Gribov-Glauber approach that treats the soft- and hard- scattering interactions differently~\cite{Roesler:2000he}. HIJING is used to study jet and the associated particle production in high energy collisions and is based on QCD-inspired models, including multiple minijet production, soft excitation, nuclear shadowing of parton distribution functions, and jet interaction in dense matter. Due to nuclear shadowing, the parton distribution functions of quarks and gluons are expected to differ from the simple superposition of their distribution in a nucleon. The initial nuclear shadowing effect on particle production is tested using two shadowing depth constant values 0.1 and 0 that represent the HIJING model predictions with and without shadowing~\cite{Gyulassy:1994ew}. PYTHIA8/Angantyr is an event generator based on the Fritiof model~\cite{Pi:1992ug}, which includes the features of multi-parton interactions and diffractive excitation in each nucleon--nucleon (NN) sub-collision. It acts as a baseline for understanding the non-collective background in the observables sensitive to collective behavior, as PYTHIA8 does not include a collective expansion stage in its description of pp collisions~\cite{Bierlich:2018xfw}. The comparison of data with the results from these phenomenological models helps to understand the relative contribution of the nuclear effects on the particle production in \pPb collisions.

For the results presented here, K$^{*}(892)^{0}$ and $\mathrm{\overline{K}}^{*}(892)^{0}$ are averaged and denoted by the symbol \kstar, while \phf is denoted by $\phim$.
 The article is organized as follows.  In Section~\ref{section:anadata}, the data sample, event and track selection criteria, the analysis techniques, the procedure of extraction of the yields, and the study of the systematic uncertainties are discussed.  In Section~\ref{section:result}, the results on the $\pt$ spectra, the d$N$/d$y$, the $\langle p_{\rm{T}}\rangle$, the $Y_\mathrm{asym}$ and  the ${Q}_\mathrm{CP}$ in $\pPb$ collisions at $\snn$ $=$ 5.02 \TeV are presented. 
 Finally, the results are summarized in Section~\ref{section:summary}.

 \section{Data analysis} \label{section:anadata}
Measurements of \kstar and \phim meson production are carried out on the data sample collected in 2016 during the second LHC run with \pPb collisions at \snn $=$ 5.02 \TeV. The resonances are reconstructed from their decay products by using the invariant-mass method. The considered decay channels are \mbox{\kstar $\rightarrow  \rm{K}^{+} \pi^{-}$} and its charge conjugate, and \phim $\rightarrow \rm{K}^{+}\rm{K}^{-}$ with respective branching ratios (BR) of \mbox{66.6 $\%$} and 49.2 $\%$~\cite{Adam:2016bpr,ALICE:2021uyz}. In the \pPb configuration, the ${}^{208}$Pb beam with energy of 1.58 \TeV per nucleon collides with a proton beam with an energy of 4 \TeV resulting in collisions at a nucleon--nucleon center-of-mass energy  \snn $=$ 5.02 \TeV~\cite{Adam:2016bpr}.
It leads to the rapidity in the center-of-mass frame being shifted by $\Delta y = -0.465$ in the direction of the  proton beam with respect to the laboratory frame. The measurements are performed in the rapidity range -1.2 $< y <$ 0.3 for five rapidity intervals with width of 0.3 units and three multiplicity classes along with a multiplicity-integrated class.
The details of the ALICE detector setup and its performance can be found in Refs.~\cite{Aamodt:2008zz,Abelev:2014ffa}. The measurements are carried out with the ALICE central barrel detectors, which are utilized for tracking, PID, and primary vertex reconstruction and are housed inside a solenoidal magnet with a magnetic field of 0.5 T. The main detectors that are used for the analyses presented here are the Inner Tracking System (ITS)~\cite{Aamodt:2010aa}, the Time Projection Chamber (TPC)~\cite{Alme:2010ke}, and the TOF (Time-Of-Flight)~\cite{DeGruttola:2014eti} detectors. These detectors have full azimuthal coverage and have a common pseudorapidity coverage of $|\eta|<$ 0.9. 
\subsection{Event and track selection and particle identification}
The trigger and event selection criteria are the same as those discussed in previous publications~\cite{Adam:2016bpr,ALICE:2021uyz}.
The events are selected with a minimum-bias trigger based on the coincidence of signals in two arrays of 32 scintillator detectors covering full azimuth and  the pseudorapidity regions 2.8 $<$ $\eta$ $<$ 5.1  (V0A) and \mbox{--3.7 $<$ $\eta$ $<$ --1.7 (V0C)}~\cite{Abbas:2013taa}. The primary vertex of the collision is determined using the charged tracks reconstructed in the ITS and the TPC. Events are selected whose reconstructed primary vertex position lies within $\pm$10 cm from the center of the detector along the beam direction. The Silicon Pixel Detector (SPD) which is the innermost detector of the ITS, is used to reject events in which multiple collision vertices are found (pile-up)~~\cite{Aamodt:2010aa}.
 In this work, approximately 540 million events are selected with the criteria described above.
The minimum-bias events are further divided into three multiplicity classes, which are expressed in percentiles according to the total charge deposited in the V0A detector~\cite{Abbas:2013taa}. The yield of \kstar and \phim mesons is measured in five rapidity regions \mbox{-1.2 $< y <$ -0.9, -0.9 $< y <$ -0.6, -0.6 $< y <$ -0.3, -0.3 $< y <$ 0 and 0 $< y <$ 0.3}  for the multiplicity classes 0--10$\%$, 10--40$\%$, 40--100$\%$ in addition to the multiplicity-integrated (0--100$\%$) measurement, corresponding to all minimum-bias events. The 10$\%$ of the events with the highest multiplicity of charged particles correspond to the 0--10$\%$ class and similarly, the  40--100$\%$ class corresponds to the lowest multiplicity. The $\langle N_\mathrm{coll} \rangle$ values are estimated from a Glauber model analysis~\cite{Miller:2007ri} of the charged particle multiplicity distribution in the V0A detector,
and they are 13.8 $\pm$ 3.8, 10.5 $\pm$ 3.9 and 4.0 $\pm$ 2.6, respectively for 0--10$\%$, 10--40$\%$, and 40--100$\%$ multiplicity classes taken from Ref.~\cite{ALICE:2014xsp}.
 Charged-particle tracks reconstructed in the TPC with $\pt>$ 0.15 \GeV/$c$ and pseudorapidity $|\eta|<$ 0.8 are selected for the analysis. The selected charged tracks should have crossed at least 70 out of 159 readout-pad rows of the TPC. The distance of closest approach of the track to the primary vertex in the longitudinal direction (DCA$_{z}$) is required to be less than 2 cm.  In the transverse plane (xy) a \pt-dependent selection of DCA$_{xy}(\pt)< $ 0.0105 + 0.035 $\pt^{-1.1}$  cm is applied.
The \kstar and \phim mesons are reconstructed from their decay daughters (pions and kaons), which are identified by measuring the specific ionization energy loss (d$E$/d$x$) in the TPC ~\cite{Alme:2010ke} and their time-of-flight information using the TOF ~\cite{DeGruttola:2014eti}.  For the selection of pions and kaons, the  measured $\langle\dEdx\rangle$ is required to be within $n\sigma_{\mathrm{TPC}}$ from the expected $\langle\dEdx\rangle$ values for a given mass hypothesis, where $\sigma_{\mathrm{TPC}}$ is the TPC $\langle\dEdx\rangle$ resolution. The values of $n$ are momentum-dependent ($p$) and are set to 6$\sigma_{\mathrm{TPC}}$, 3$\sigma_{\mathrm{TPC}}$, and 2$\sigma_{\mathrm{TPC}}$ in the momentum intervals \mbox{$p$ $<$ 0.3 \GeV/$c$}, \mbox{0.3 $<$ $p$ $<$ 0.5 \GeV/$c$} and $p$ $>$ 0.5 \GeV/$c$, respectively.
If the TOF information is available for the considered tracks, it is used for pion and kaon identification in addition to the TPC one by requiring the time-of-flight of the particle to be within 3$\sigma_{\mathrm{TOF}}$ from the expected value for the considered mass hypothesis, where $\sigma_{\mathrm{TOF}}$ is the time-of-flight resolution of the TOF.

\subsection{Yield extraction}
The \kstar and \phim resonances are reconstructed from their decay products using the invariant-mass reconstruction technique described in Refs.~\cite{Adam:2016bpr,ALICE:2021uyz}. The invariant-mass distributions of K$^{\pm}\pi^{\mp}$ and K$^+$K$^-$ pairs in the same event are reconstructed. The shape of uncorrelated background is estimated using two techniques, namely mixed-event and like-sign methods. In the mixed-event method, the shape of the uncorrelated-background distribution for \kstar (\phim ) is obtained by combining pions (kaons) from a given event with opposite-sign kaons from other events. Each event is mixed with five different events to reduce the statistical uncertainties of the estimated uncorrelated-background distribution. The events which are mixed are required to have similar characteristics, i.e.\ the longitudinal position of the primary vertices should differ by less than 1 cm, and the multiplicity percentiles, computed from the V0A amplitude, should differ by less than 5$\%$. The mixed-event distributions for \kstar (\phim) candidates are normalized in the invariant mass interval \mbox{1.1 $ < M_\mathrm{K\pi} <$ 1.15 \GeV/$c^{2}$} (\mbox{1.06 $< M_\mathrm{KK} <$ 1.09 \GeV/$c^{2}$}), which is well separated from signal peak.
\begin{figure}[H]
	\begin{center}
		\includegraphics[width =0.8\textwidth]{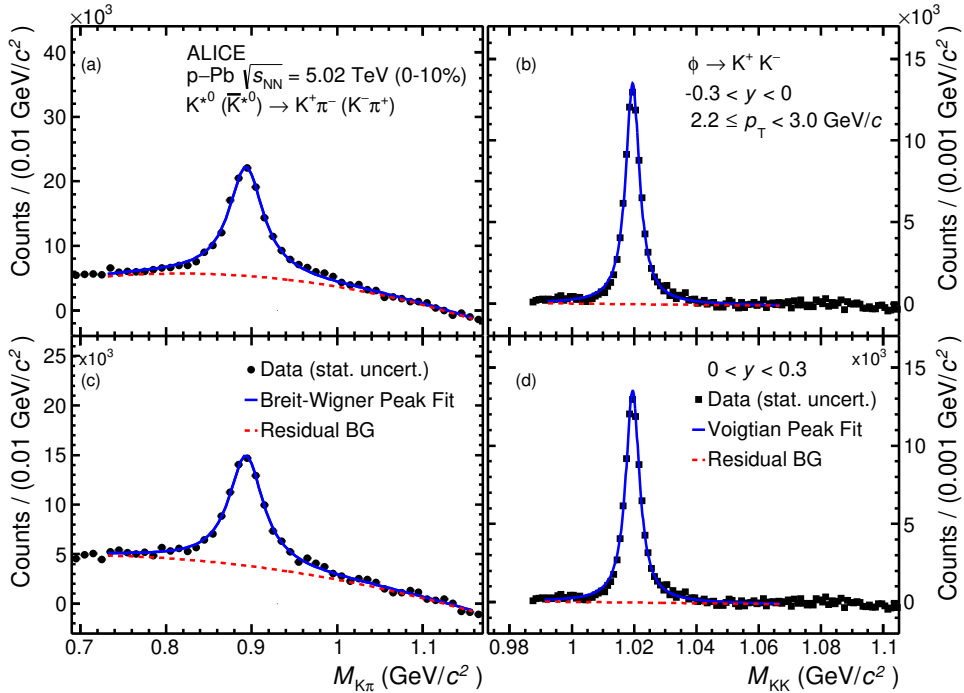}
	\end{center}
	\caption{Invariant-mass distributions after combinatorial background subtraction for \kstar and \phim candidates in the multiplicity class 0--10$\%$ and  transverse momentum range 2.2 $\leq$ $p_{\mathrm{T}}$  $<$ 3.0 \GeV/$c$ in the rapidity interval \mbox{-0.3 $<  y <$ 0} (panels (a) and (b))  and 0 $<  y <$ 0.3 (panels (c) and (d)). The \kstar peak is described by a Breit-Wigner function whereas the \phim peak is fitted with a Voigtian function. The residual background is described by a polynomial function of order 2.}
	\label{fig:Invmass}
\end{figure}
In the like-sign method, tracks with the same charge 
from the same event are paired to estimate the uncorrelated background contribution. The invariant-mass distribution for the uncorrelated background is obtained as the geometric mean 2$\sqrt{n^{++}\times n^{--}}$, where $n^{++}$ and $n^{--}$ are the number of positive-positive and negative-negative pairs in each invariant-mass interval, respectively.
The mixed-event technique is used as the default method to extract the yields for both \kstar and \phim mesons, while the difference with respect to the yield obtained using the combinatorial background from the like-sign method is included in the estimation of the systematic uncertainty. 
After the subtraction of the combinatorial background, the invariant-mass distribution consists of a resonance peak sitting on the top of a residual background of correlated pairs. The residual background originates from correlated pairs from jets, misidentification of pions and kaons from \kstar and \phim meson decays, and partially reconstructed decays of higher-mass particles~\cite{Adam:2016bpr}.
Figure~\ref{fig:Invmass} shows the  ${K^{\pm}\rm{\pi^{\mp} }}$ and ${\rm{K^{+}K^{-}}}$ invariant-mass distributions after subtraction of mixed-event background in the transverse
momentum interval 2.2 $\leq$ $p_{\mathrm{T}}$  $<$ 3.0 \GeV/$c$ for the rapidity intervals -0.3 $< y <$ 0 (panels (a) and (b)) and 0 $< y <$ 0.3 (panels (c) and (d)) in the 0--10$\%$ multiplicity class. 

The signal peak is fitted with a Breit-Wigner and a Voigtian function (convolution of Breit-Wigner and Gaussian functions) for $\kstar$ and $\phim$ resonances, respectively. For the  $\kstar$,  a pure Breit-Wigner is used because 
the invariant mass resolution 
is negligible with respect to the natural width of the resonance peak. A second-order polynomial function is used to describe the shape of the residual background for both resonances. The fit to the invariant-mass distribution is performed in the interval 0.75 $< M_{\rm{K}\pi} <$ 1.15 \GeV/$c^{2}$ (0.99 $< M_\mathrm{KK} <$ 1.07 \GeV/$c^{2}$) for $\kstar$($\phim$). The widths of $\kstar$ and $\phim$ peaks are fixed to their known widths $\Gamma$(K*$^{0}$) $=$ 47.4 $\pm$ 0.6 MeV/$c^{2}$, $\Gamma$($\phi$) $=$ 4.26 $\pm$ 0.04 MeV/$c^{2}$~\cite{Tanabashi:2018oca}, whereas the resolution parameter of the Voigtian function for \phim is kept as a free parameter.
In the estimation of the systematic uncertainties, the width of the Breit-Wigner is taken as a free parameter.
The mass and width values extracted from the fit have similar magnitude and trend with \pt as reported in previous publications~\cite{Adam:2016bpr,ALICE:2021uyz,Acharya:2019wyb,Acharya:2019bli}. In the present study, it is found that the mass and width obtained from the fit are independent of rapidity and multiplicity for both $\kstar$ and $\phim$ mesons. 
 The sensitivity of the systematic uncertainty to the choice of the fitting range, normalization interval of the mixed-event background, shape of the residual background function, width, and resolution parameters have been studied by varying the fit configuration, as described in \mbox{Section~\ref{subsection:sys}}. The raw yields of $\kstar$ and $\phim$ mesons are extracted in the transverse momentum range from \mbox{0.8 to 16 \GeV/$c$} for various rapidity intervals and multiplicity classes. \\
 To obtain the transverse momentum spectra, the raw yields are normalized by the number of accepted non-single-diffractive (NSD) events and corrected for the branching ratio and the detector acceptance ($A$) times the reconstruction efficiency ($\epsilon_{\rm rec }$). The $A \times \epsilon_{\rm rec}$ is obtained from a Monte Carlo (MC) simulation based on the DPMJET~\cite{Roesler:2000he} event generator and the GEANT3 package to model the transport of the generated particles through the ALICE detector~\cite{Brun:1119728}. 
 The $A \times \epsilon_{\rm rec}$ is defined as the ratio of the reconstructed \pt spectra of $\kstar$ ($\phi$) mesons in a given rapidity interval to the generated ones in the same rapidity interval. 
 The track and PID selection criteria applied to the decay products of resonances in the MC are identical to those used in the data. Since the efficiency depends on \pt and the \pt distributions of $\kstar$ and $\phi$ mesons from DPMJET are different from the real data, a re-weighting procedure is applied to match the generated \pt shapes to the measured ones.
 
The effect of the re-weighting on $A\times\epsilon_\mathrm{rec}$ depends on $\pt$ and amounts to $\sim$5--17$\%$ at $\pt$ $<1.5$~\GeV/$c$. At higher $\pt$, the effect is negligible.
The effect of re-weighting also depends on rapidity at low \pt. The re-weighted $A \times \epsilon_\mathrm{rec}$ is used to correct the raw $\pt$ distribution. The $A \times \epsilon_\mathrm{rec}$ is calculated for each rapidity interval and multiplicity class considered in the analysis. The $A \times \epsilon_\mathrm{rec}$ as a function of \pt  shows a rapidity dependence for a given multiplicity class, however, no significant multiplicity dependence of $A \times \epsilon_\mathrm{rec}$ is observed for a given rapidity interval.

\subsection{Systematic uncertainties} \label{subsection:sys}
The procedure to estimate the systematic uncertainties is similar to the one adopted in previous analyses~\cite{Adam:2016bpr,ALICE:2021uyz}. The sources of systematic uncertainties on the measured yield of \kstar and \phim mesons are signal extraction, track selection criteria, PID, global tracking efficiency, uncertainty on the material budget of the ALICE detector, and the hadronic interaction cross section in the detector material. A summary of the systematic uncertainties on the \pt spectra is given in Table~\ref{t1}. The uncertainty due to signal extraction is estimated from the variation of the yields when varying the invariant mass fit range, the treatment of the Breit-Wigner width in the fits, the mixed-event background normalization interval, the choice of residual background function, and the method to determine the combinatorial background. The fitting range is varied by 50 MeV$/c^{2}$ for \kstar and 10 MeV$/c^{2}$ for \phim. The normalization interval of the mixed-event background is varied by 150 (50) MeV$/c^{2}$ with respect to the default value for \kstar(\phim). The width of \kstar and \phim resonances is left as a free parameter in the fit, instead of fixing it to the world-average value. For $\phim$ resonances, the effect on the yield due to the variation of resolution parameter ($\sigma$ of the Gaussian function in the Voigtian distribution) is also considered. The residual background is parameterized using first-order and third-order polynomial functions for estimating its contributions to the systematic uncertainties. The combinatorial background from the like-sign method is used instead of the one from event mixing. 
The estimated systematic uncertainties due to the yield extraction is \mbox{5.2$\%$} for $\kstar$ and 3.3$\%$ for $\phim$. The systematic effects due to charged track selection have been studied by varying the selection criteria on the number of crossed rows in the TPC,  the ratio of the numbers of TPC crossed rows to findable clusters, and the DCA to the primary vertex of the collisions. The estimated uncertainties due to the track selection is 2.5$\%$ for $\kstar$ and about 5$\%$ for the $\phim$ mesons. To estimate the systematic uncertainty due to the PID, the selections on the d$E/$d$x$ and time-of-flight of the pions and kaons are varied. 
Two momentum-independent selections: 2$\sigma_{\rm{TPC}}$ with  3$\sigma_{\rm{TOF}}$  and 2$\sigma_{\rm{TPC}}$ only, are used for both \kstar and \phim.
The estimated systematic uncertainties are 3$\%$ for $\kstar$ and 1.7$\%$  for $\phim$. The uncertainty due to the global tracking efficiency, description of the detector material budget in the simulation, and the cross sections for hadronic interactions in the material are taken from Ref.~\cite{Adam:2016bpr}. The total systematic uncertainty is taken as the quadratic sum of all contributions leading to 7.5$\%$ for $\kstar$ and 7.3$\%$ for $\phim$ mesons. No multiplicity and rapidity dependence of the systematic uncertainties is observed. Therefore, the systematic uncertainties on the $\pt$ spectra determined for  minimum-bias events in the rapidity interval 0 $< y <$ 0.3 are assigned in all rapidity intervals and multiplicity classes.
  \begin{table}
	\begin{center}
	\caption{Relative systematic uncertainties for \kstar and \phim yields 
		in  \pPb collisions at \snn $=$ 5.02 \TeV. The quoted relative uncertainties are averaged over $\pt$ in the range 0.8--16 \GeV/$c$. 
		The total systematic uncertainty is the sum in quadrature of the uncertainties due to each source. }
			\label{t1}
			\scalebox{0.8}
				{
					\begin{tabular}{ccccc}
						\hline \hline
						Systematic uncertainty     &  \multicolumn{1}{c} ~ \kstar  &  \multicolumn{2}{c}~\phim~~ \\
						\hline                                                              
						Yield extraction ($\%$)              & 5.2                         & 3.3 \\
						Track selection ($\%$)               & 2.5                         & 5.0  \\
						Particle identification ($\%$)       & 3.0                         &1.7 \\
						Global tracking efficiency ($\%$)    & 3.0                          & 2.1 \\
     					Material budget ($\%$)               & 1.2                          &2.2 \\
    					Hadronic Interaction ($\%$)          & 1.9                          &2.4   \\
    					\hline
    					 Total ($\%$)                        & 7.5                          & 7.3 \\
						\hline \hline
					\end{tabular}
				}
		\end{center}
	\end{table}

The systematic uncertainties on $Y_{\rm{asym}}$ are estimated by considering the same approaches and variations as for the corrected yields. 
The systematic uncertainties due to signal extraction and PID are uncorrelated among different rapidity intervals whereas the other sources of systematic uncertainties such as track selections, global tracking uncertainties, material budget and hadronic interactions are correlated and cancel out in the $Y_{\rm{asym}}$ ratio.
For the uncorrelated sources of uncertainty, the same variations considered for the yields were studied by estimating their effects on the $Y_{\rm{asym}}$ ratio. The resulting uncertainty was estimated to be about 2.5$\%$ (2$\%$)  for \kstar ($\phim)$ mesons. 
No multiplicity and rapidity dependence of the  uncertainties is observed for $Y_{\rm{asym}}$. Therefore, the systematic uncertainties determined for minimum-bias events are assigned to the ratios in the different rapidity intervals and multiplicity classes.
The systematic uncertainties on the ratios (d$N/$d$y$)/(d$N/$d$y)_{y=0}$ and $\langle\pt\rangle$/$\langle\pt\rangle_{y=0}$ as a function of rapidity are calculated in a similar way as for $Y_{\rm{asym}}$.
The systematic uncertainties on ( d$N/$\rm{d}$y$)/(d$N/$\rm{d}$y)_{y=0}$ and  $\langle\pt\rangle$/$\langle\pt\rangle_{y=0}$ are $~$2.2$\%$ (2$\%$) and $~$1.2$\% $  (1$\%$) for \kstar ($\phim)$, respectively.

\section{Results and discussion} \label{section:result}
The rapidity and multiplicity dependence results on the $\pt$ spectra, the d$N$/d$y$, the $\langle\pt\rangle$, the $Y_\mathrm{asym}$, and the ${Q}_\mathrm{CP}$ in $\pPb$ collisions at $\snn$ $=$ 5.02 \TeV are discussed. The measurements are also compared with various model predictions. 
\subsection{Transverse momentum spectra}
\begin{figure}[H]
   \begin{center}
 \includegraphics[width = 0.8\textwidth]{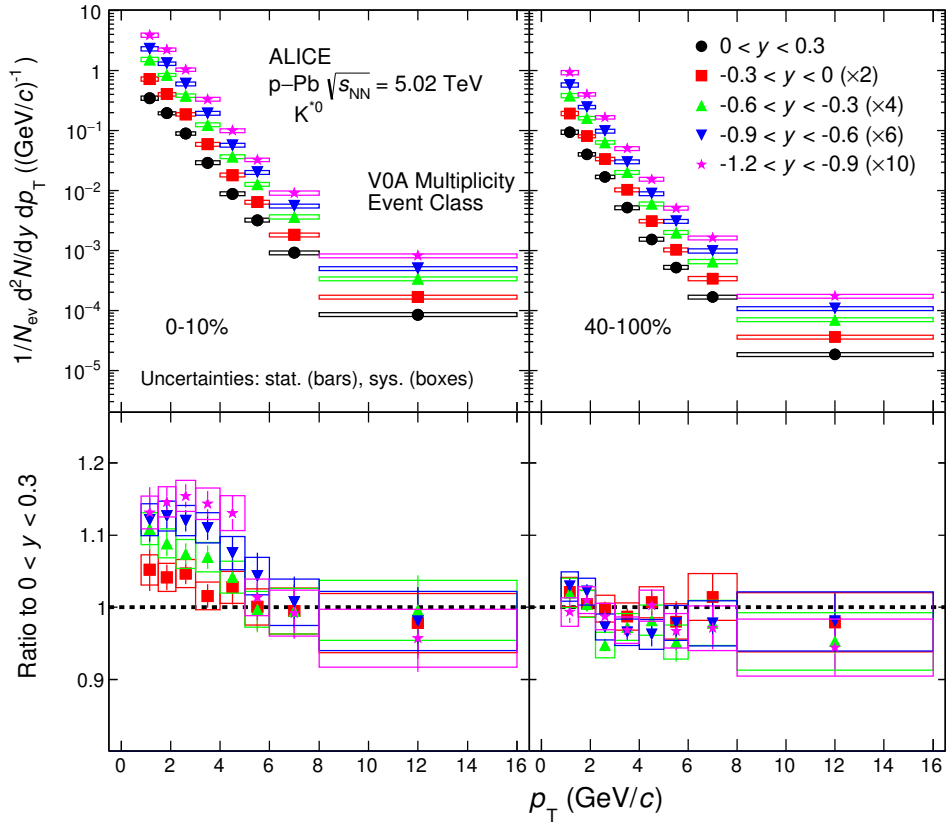}
   \end{center}
 \caption{ Top panels: The transverse momentum spectra of \kstar for five rapidity intervals within $-$1.2 $< y < $ 0.3 and for two multiplicity classes (0--10$\%$, 40--100$\%$) in \pPb collisions  at \mbox{\snn = 5.02 \TeV}. The data for different rapidity intervals are scaled for better visibility. Bottom panels: The ratios of the \pt spectra in various rapidity intervals to that in the interval 0 $< y < $ 0.3 for a given multiplicity class. The statistical and systematic uncertainties are shown as bars and boxes around the data points, respectively.}   
  \label{fig:rapspectrakstar}
  \end{figure}

\begin{figure}[H]
   \begin{center}
 \includegraphics[width = 0.8\textwidth]{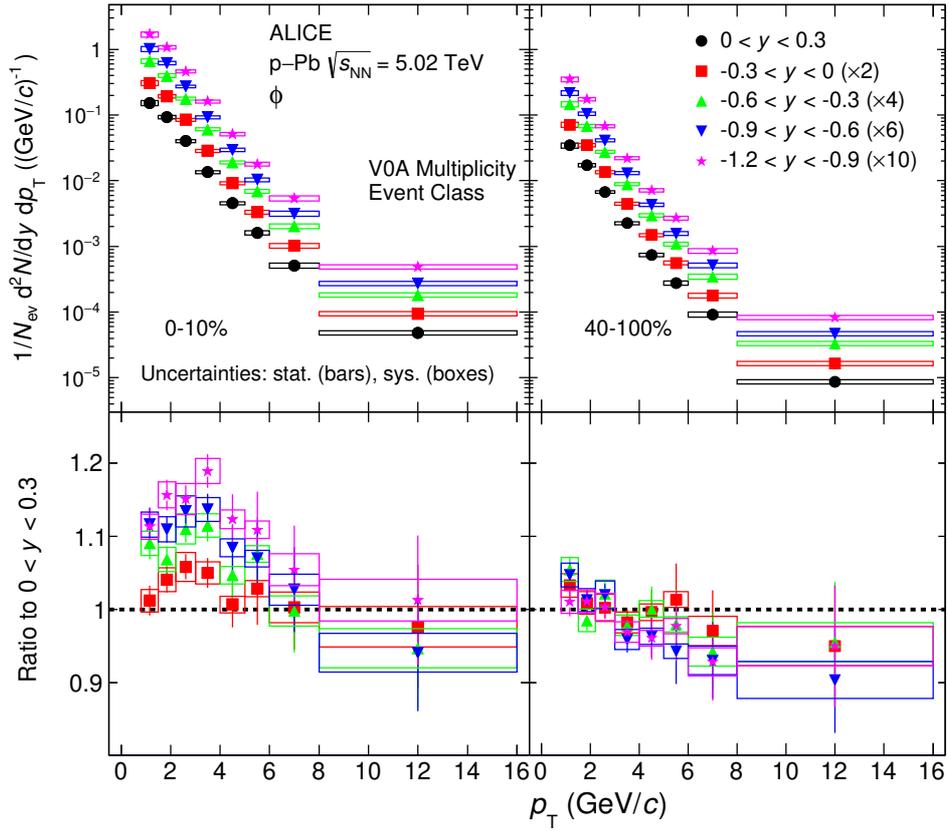}
   \end{center}
    \caption{ Top panels: The transverse momentum spectra of \phim for five rapidity intervals within $-$1.2 $< y < $ 0.3 and for two multiplicity classes (0--10$\%$, 40--100$\%$) in \pPb collisions at \mbox{\snn = 5.02 \TeV}. The data for different rapidity intervals are scaled for better visibility. Bottom panels: The ratios of the \pt spectra in various rapidity intervals to that in the interval 0 $< y < $ 0.3 for a given multiplicity class. The statistical and systematic uncertainties are shown as bars and boxes around the data points, respectively.}    
 \label{fig:rapspectraphi}
  \end{figure}

Figure~\ref{fig:rapspectrakstar} and Fig.~\ref{fig:rapspectraphi} show the \pt spectra of \kstar and \phim mesons in \pPb collisions at \snn $=$ 5.02 \TeV for five rapidity intervals within -1.2 $< y < $ 0.3 and for two multiplicity classes \mbox{0--10$\%$ and 40--100$\%$}, respectively.
The ratios of the \pt spectra in different rapidity intervals to that in the interval \mbox{0 $< y < $ 0.3} are presented in the bottom panels of Fig.~\ref{fig:rapspectrakstar} and  Fig.~\ref{fig:rapspectraphi}.
The measured \pt spectra of \kstar and \phim mesons in the 0--10$\%$ multiplicity class show a rapidity dependence at low $\pt$ ($<$ 5 \GeV/$c$) indicating that the production of these resonances is higher in the Pb-going direction ($y <  $ 0) than in the p-going direction ($y > $ 0).
 For high \pt, no rapidity dependences are observed. 
  
  \subsection{Integrated particle yield and mean transverse momentum}
\begin{figure}[H]
 \begin{center}
    \includegraphics[scale=0.7]{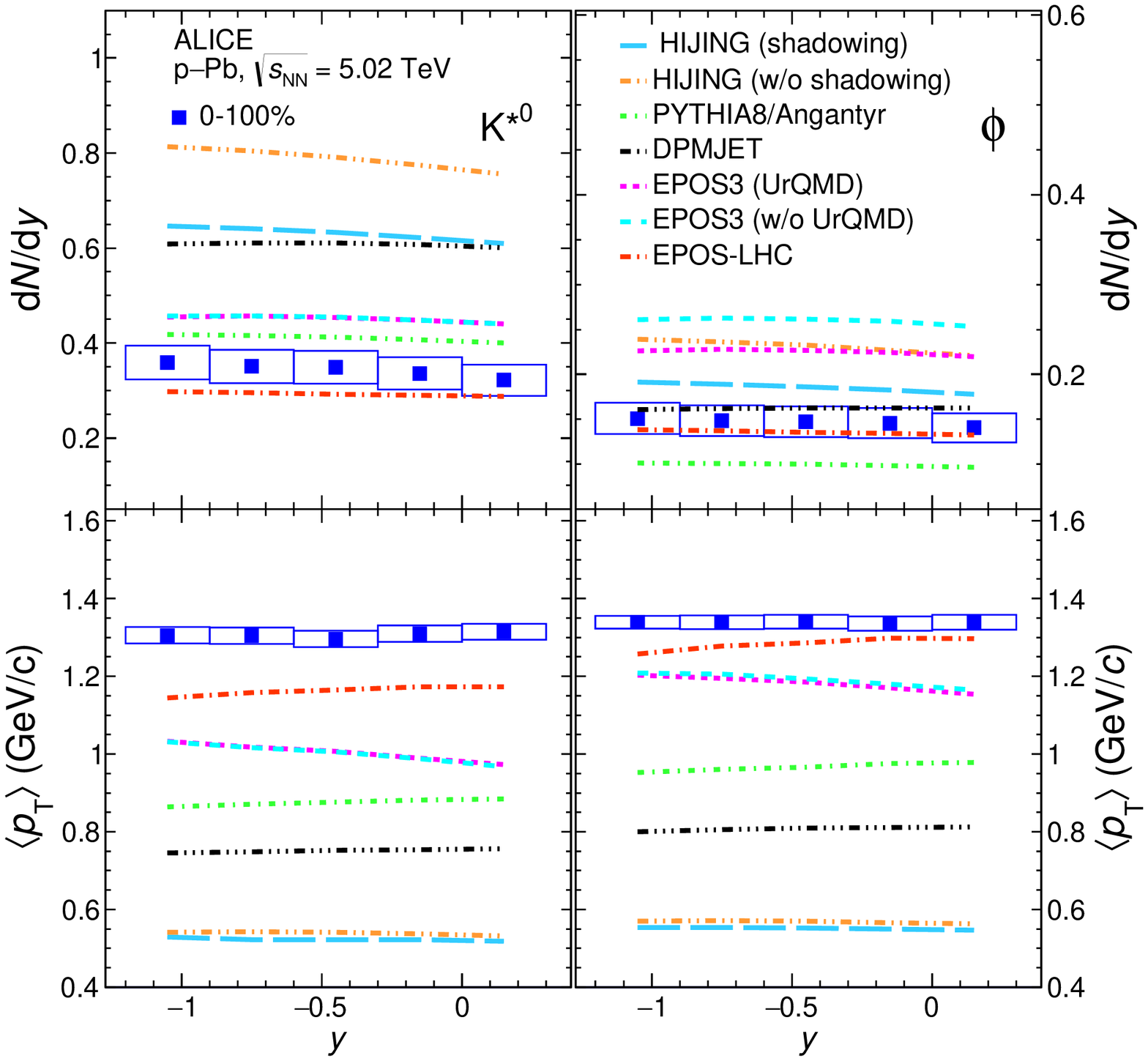}          
	\end{center}
   \caption{The \pt integrated yield (d$N/\mathrm{d}y$) (top panels) and mean transverse momentum ($\langle p_{\mathrm{T}} \rangle$) (bottom panels) for \kstar (left) and \phim (right) mesons as a function of $y$
   	 measured for the multiplicity class 0-100$\%$ in \pPb collisions at  \snn $=$ 5.02 \TeV. 
   	 The predictions from EPOS-LHC~\cite{Pierog:2013ria},  EPOS3 with and without UrQMD~\cite{Werner:2013tya,Knospe:2021jgt}, DPMJET~\cite{Roesler:2000he}, HIJING~\cite{Gyulassy:1994ew}, and PYTHIA8/Angantyr~\cite{Bierlich:2018xfw} are shown as different curves. The statistical uncertainties are represented as bars whereas the boxes indicate total systematic uncertainties. }
  	\label{fig:mdndymeanpt}
 \end{figure}

 The d$N/$\rm{d}$y$ and the $\langle\pt\rangle$ are obtained from the transverse momentum spectra in the measured \pt interval and using a fit function to account for the contribution of \kstar and \phim mesons in unmeasured regions.
 The spectra are fitted with a L\'evy-Tsallis function~\cite{Tsallis:1987eu} and the fit function is extrapolated to  unmeasured regions at low \pt ( $<$ 0.8 \GeV/$c$). The integral of the fit function in the extrapolated region accounts for 33$\%$ (39$\%$) of the total yield in the 0--10$\%$ (40--100$\%$) multiplicity class for both \kstar and \phim mesons. The contribution of the extrapolated yield at low \pt  is the same for all rapidity intervals.
 The contribution of the yield in the unmeasured region at high \pt ( $>$ 16 \GeV/$c$) is negligible for both \kstar and \phim mesons.
 The extrapolated yield contribution at low $\pt$ obtained with different fitting functions (i.e.,  $m_\mathrm{T}$-exponential, Bose-Einstein and Boltzmann-Gibbs Blast-Wave function~\cite{ALICE:2020jsh}) and that obtained with the default L\'evy-Tsallis function is 5$\%$ (8$\%$) for the 0--10$\%$ (40--100$\%$) included as the systematic uncertainties in the d$N/$\rm{d}$y$ and it varies by 2--5 $\%$ for the $\langle p_{\mathrm{T}} \rangle$. 
 In Fig.~\ref{fig:mdndymeanpt} the d$N/$\rm{d}$y$ (top panels) and $\langle\pt\rangle$ (bottom panels) of \kstar (left) and \phim (right) mesons are shown as a function of $y$ for minimum-bias \pPb collisions at \mbox{$\snn$ = 5.02 \TeV}.
 The central values of the d$N/$\rm{d}$y$ of both \kstar and \phim mesons decrease slightly from the rapidity interval $-$1.2 $< y  <$ $-$0.9 to 0 $< y <$ 0.3  even though within the systematic uncertainties all the data points are compatible among each other. Nevertheless, considering that the systematic uncertainties are mostly correlated among the rapidity intervals, the measured d$N/$\rm{d}$y$ values suggest a decreasing trend with increasing $y$ in the rapidity interval covered by the measurement. The $\langle\pt\rangle$ is constant as a function of rapidity for both \kstar and \phim resonances.
The model predictions from EPOS-LHC~\cite{Pierog:2013ria}, EPOS3 with and without UrQMD ~\cite{Werner:2013tya,Knospe:2021jgt}, DPMJET~\cite{Roesler:2000he}, HIJING~\cite{Gyulassy:1994ew}, and 
PYTHIA8/Angantyr~\cite{Bierlich:2018xfw} are also shown in the Fig.~\ref{fig:mdndymeanpt}. 
In general, the models show a similar trend with rapidity as the data except EPOS3 with and without UrQMD for $\langle\pt\rangle$, which shows a pronounced decreasing trend with rapidity. 
All the model predictions shown in Fig.~\ref{fig:mdndymeanpt} underestimate the  $\langle\pt\rangle$
of both meson species. For the d$N/$\rm{d}$y$, HIJING and EPOS3 with and without UrQMD overpredict the measured values for both \kstar and \phim, while PYTHIA8/Angantyr overpredicts the \kstar and underpredicts the \phim yield. EPOS-LHC provides the best overall description of the d$N/$\rm{d}$y$ and  $\langle\pt\rangle$ measurements for \kstar and \phim mesons.
The  $\langle\pt\rangle$ also shows a flat behavior as a function of rapidity for all the considered multiplicity classes as it can be seen in Fig.~\ref{fig:dndymeanptmult} of Appendix~\ref{section:appendix}. 

\begin{figure}[H]
  	\begin{center}
  		\includegraphics[scale=0.7]{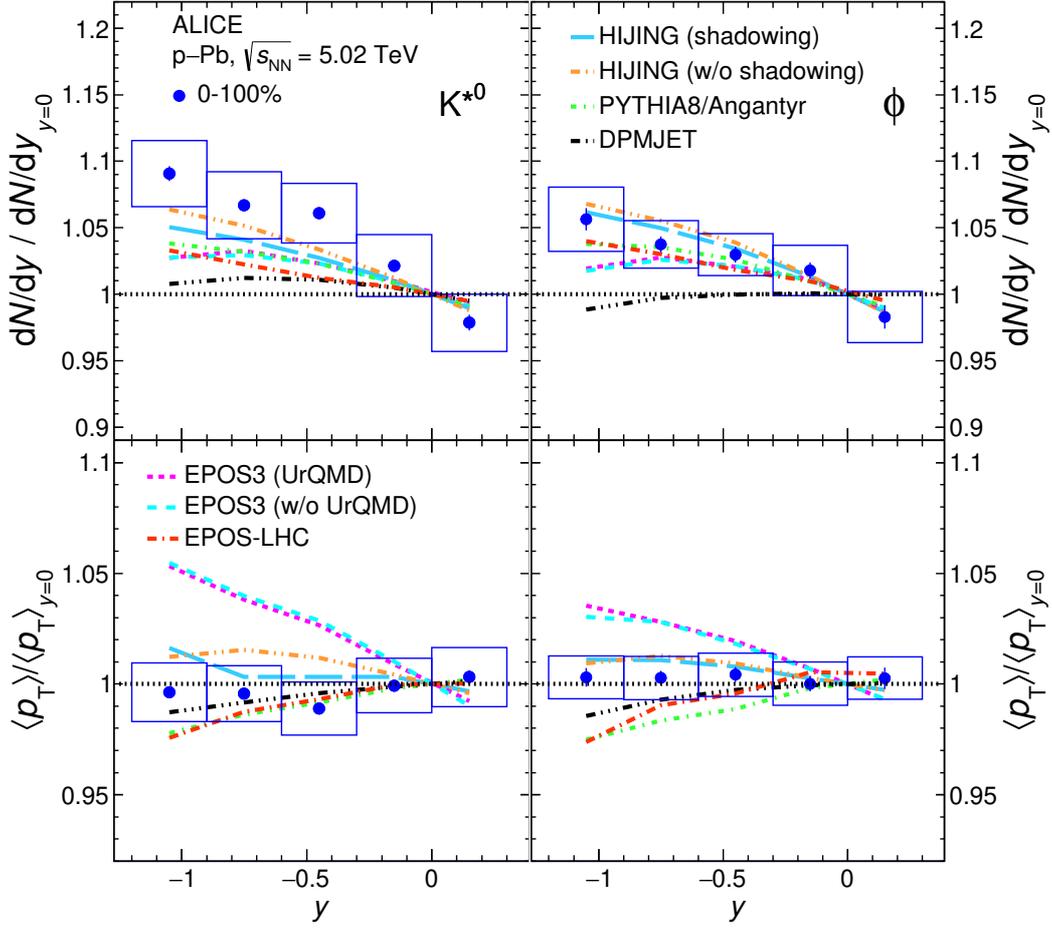}          
  	\end{center}
  	\caption{The \pt integrated yield (d$N/\mathrm{d}y$) (upper panels) and mean transverse momentum $\left(\langle\pt\rangle\right)$ (bottom panels) for \kstar (left) and \phim (right) mesons as a function of $y$,  divided by  the d$N/\mathrm{d}y$ and $\langle p_{\mathrm{T}} \rangle$  at $y = 0$ for the multiplicity class 0--100$\%$ in \pPb collisions at  \snn $=$ 5.02 \TeV.
  		The predictions from EPOS-LHC~\cite{Pierog:2013ria}, EPOS3 with and without UrQMD~\cite{Werner:2013tya,Knospe:2021jgt}, DPMJET~\cite{Roesler:2000he}, HIJING~\cite{Gyulassy:1994ew}, and PYTHIA8/Angantyr~\cite{Bierlich:2018xfw} are shown as different curves.
  		The statistical uncertainties are represented as bars whereas the boxes indicate total systematic uncertainties. }
  	\label{fig:rdndymeanpt}
  \end{figure}

A similar behavior in the average transverse kinetic energy as a function of rapidity for strange hadrons was reported in Ref.~\cite{CMS:2016zzh}. The rapidity dependence of d$N/$\rm{d}$y$ and $\langle\pt\rangle$ for \kstar and \phim mesons in the multiplicity class 0--100$\%$ is further studied by dividing the d$N/$\rm{d}$y$ and $\langle\pt\rangle$ values in a given rapidity interval by the corresponding values at $y=0$, as shown in Fig.~\ref{fig:rdndymeanpt}.  The d$N/$\rm{d}$y$ and  $\langle\pt\rangle$ value at $y=0$
  is computed from the $\pt$ spectrum measured in the rapidity interval -0.3$ < y < $0.3.
  The systematic uncertainties on these ratios are estimated by studying the effects of the variations directly on the ratios as discussed in Section~\ref{subsection:sys}. This procedure takes into account the correlation of the systematic uncertainties across rapidity bins: as a result, these ratios have smaller systematic uncertainties than those on the d$N/$\rm{d}$y$ and $\langle\pt\rangle$, and allow for a better insight into the $y$ dependence. The ratio (d$N/$\rm{d}$y$)/(d$N/$\rm{d}$y)_{y=0}$ decreases with rapidity, whereas $\langle p_{\mathrm{T}} \rangle$/$\langle p_{\mathrm{T}}\rangle _{y=0}$ shows a flat behavior within uncertainties as a function of rapidity for \kstar and \phim mesons. The measurements are compared with various model predictions.
  The predictions from HIJING qualitatively reproduce the trend and are the closest to the data for both \kstar and \phim.  The predictions from PYTHIA8/Angantyr, DPMJET, EPOS-LHC, EPOS3 with and without UrQMD show a decreasing trend of (d$N/$\rm{d}$y$)/(d$N/$\rm{d}$y)_{y=0}$ with increasing $y$, 
  but the rapidity dependence is less pronounced than the one in data, as it can be seen by the fact that they all tend to underestimate the measured yield ratios in the lowest rapidity intervals, especially for \kstar meson.  

 For $\langle p_{\mathrm{T}} \rangle$/$\langle p_{\mathrm{T}}\rangle_{y=0}$ as a function of $y$, also shown in Fig.~\ref{fig:rdndymeanpt}, EPOS3 with and without UrQMD overestimate the measurements at low $y$ and predict a marked decreasing trend of $\langle p_{\mathrm{T}} \rangle$/$\langle p_{\mathrm{T}}\rangle _{y=0}$ with rapidity, which is not supported by the data. From the other models, less pronounced trends are expected, which are consistent with the data. In particular, HIJING predicts a slightly decreasing \mbox{$\langle p_{\mathrm{T}} \rangle$/$\langle p_{\mathrm{T}}\rangle _{y=0}$} with increasing rapidity, while PYTHIA8/Angantyr, DPMJET, and EPOS-LHC predict a slightly increasing trend.
 
 Similar studies of the ratio $\langle p_{\mathrm{T}} \rangle$/$\langle p_{\mathrm{T}}\rangle _{y=0}$ of charged hadrons in \pPb collisions at \snn = 5.02 \TeV compared with the predictions of hydrodynamics and color-glass condensate (CGC) model were reported in~\cite{Bozek:2013sda}. Predictions from hydrodynamic calculations show a decrease in $\langle\pt\rangle$ with rapidity, whereas CGC predicts an increase in $\langle\pt\rangle$ with rapidity~\cite{Bozek:2013sda}, while the data are flat within uncertainties. The d$N/$\rm{d}$y$ and  $\langle\pt\rangle$ increase with multiplicity at midrapidity as observed 
 for light-flavor hadrons and resonances in pp and \pPb collisions\cite{ALICE:2020jsh,Adam:2016bpr, ALICE:2021uyz}. A similar behavior is observed in this article for \kstar and \phim in all the different rapidity intervals shown 
 in Fig.\ref{fig:dndymeanptmult} in Appendix~\ref{section:appendix}. 
 
  \subsection{Rapidity asymmetry}
\begin{figure}[H]
	\begin{center}
		\includegraphics[scale = 0.7]{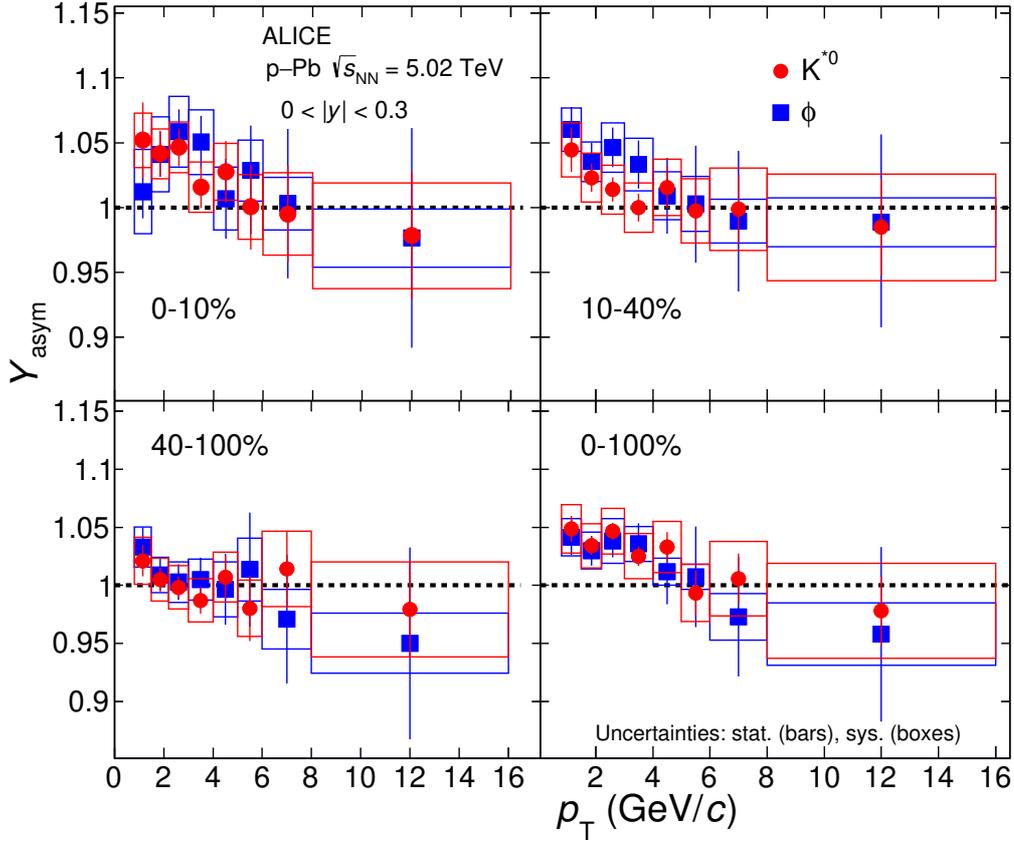}
	\end{center}
	\caption{Rapidity asymmetry ($Y_\mathrm{asym}$) of K$^{*0}$ (red circles) and $\phi$ (blue squares) meson production as a function of $p_\mathrm{T}$ in the rapidity range 0 $<$ $|y|$ $<$ 0.3 for various multiplicity classes in p--Pb collisions at \snn = 5.02 \TeV. The statistical uncertainties are shown as bars whereas the boxes represent the systematic uncertainties on the measurements.}
	\label{fig:yasymspecies}
\end{figure}

\begin{figure}[H]
	\begin{center}
		\includegraphics[scale = 0.7]{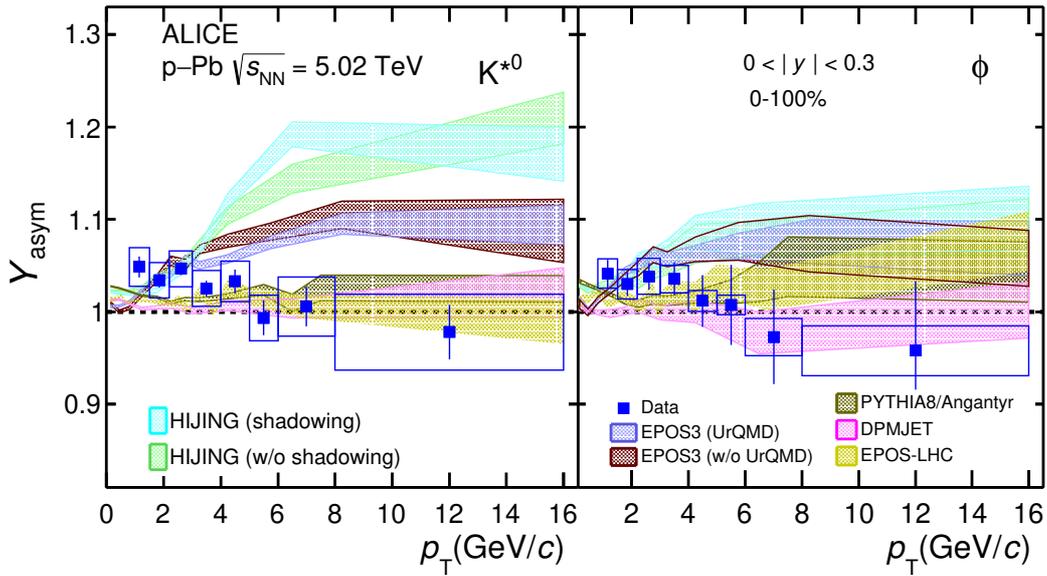}
	\end{center}
	\caption{The comparison of experimental results  of $Y_\mathrm{asym}$ for \kstar and \phim meson production as a function of $p_\mathrm{T}$ in the rapidity range 0 $< |y| <$ 0.3  with the model predictions from EPOS-LHC~\cite{Pierog:2013ria}, EPOS3 with and without UrQMD~\cite{Werner:2013tya,Knospe:2021jgt}, DPMJET~\cite{Roesler:2000he}, HIJING~\cite{Gyulassy:1994ew}, and PYTHIA8/Angantyr~\cite{Bierlich:2018xfw}.  Data points are shown with blue markers, and model predictions are shown by different color bands, where bands represent the statistical uncertainity of the model. The statistical uncertainties on the data points are represented as bars whereas the boxes indicate total systematic uncertainties.}
	\label{fig:myasym}
\end{figure}

The rapidity asymmetry ($Y_{\mathrm{asym}}$) is calculated from \kstar and \phim mesons yields in -0.3$ < y < $0 and 0$ < y < $0.3, as defined by Equation~\ref{eqn:asym}. Figure~\ref{fig:yasymspecies} shows the $Y_\mathrm{asym}$ of \kstar and $\phim$ mesons in the measured \pt intervals for various multiplicity classes in $\pPb$ collisions at \snn = 5.02 \TeV.
The $Y_{\mathrm{asym}}$ values for \kstar and $\phim$  as a function of \pt are consistent within uncertainties for all multiplicity classes. The $Y_\mathrm{asym}$ values deviate from unity at low $\pt$ ( $<$ 5 \GeV/$c$), suggesting the presence of a rapidity dependence in the nuclear effects. The deviations are more significant for events with high multiplicity. The $Y_\mathrm{asym}$ values are consistent with unity at high $\pt$ ( $>$ 5 \GeV/$c$) for all multiplicity classes, suggesting  the absence of nuclear effects at high $\pt$ for the production of \kstar and $\phim$ mesons in \pPb collisions. Similar results have been reported for charged hadrons, pions, protons in d--Au collisions at \mbox{\snn = 200 \GeV} by the STAR Collaboration~\cite{STAR:2006kxj} and for charged hadrons and multi-strange hadrons in \pPb collisions at \snn = 5.02 \TeV by the CMS Collaboration as discussed in Refs.~\cite{CMS:2016zzh,CMS:2019isl}.
Figure~\ref{fig:myasym} shows the comparison of the measured $Y_\mathrm{asym}$ for \kstar and \phim mesons as a function of \pt
in minimum-bias events (0-100$\%$) with the model predictions from EPOS-LHC, HIJING with and without shadowing, DPMJET, PYTHIA8/Angantyr, and EPOS3 with and without UrQMD. 
HIJING with and without shadowing, and EPOS3 with and without UrQMD describe the measured $Y_\mathrm{asym}$ at low \pt within uncertainties, but they significantly overestimate the data at high \pt , predicting an increasing trend with \pt (more pronounced for \kstar than for \phim) that is not supported by the measurements, which are consistent with a flat or decreasing trend for both meson species. Model predictions from EPOS-LHC, PYTHIA8/Angantyr, and DPMJET for \kstar and DPMJET for $\phim$ at high \pt  are in agreement with the data within uncertainties. 

\subsection{Nuclear modification factor}
\begin{figure}[H]
	\begin{center}
	\includegraphics[scale = 0.8]{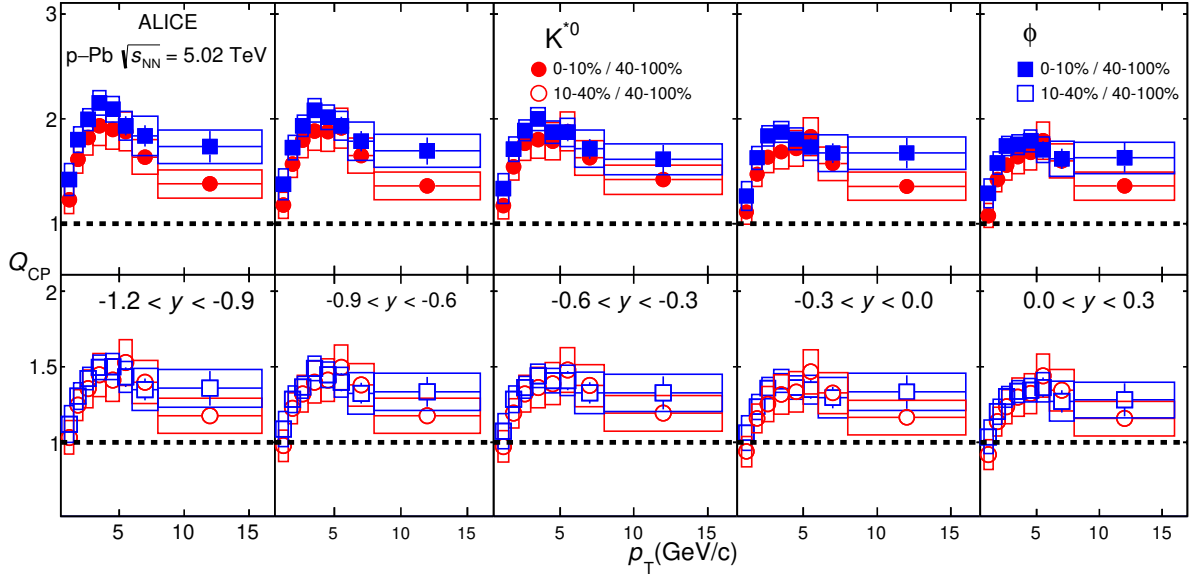}
	\end{center}
	\caption{The ${Q}_\mathrm{CP}$ of  \kstar(red  circles) and \phim(blue squares) mesons as a function of $\pt$ for 0--10$\%$ / 40--100$\%$ (top panels) and 10--40$\%$ / 40--100$\%$ (bottom panels) in various rapidity intervals within the range $-$1.2 $<$ $y$ $<$ 0.3 in $\pPb$ collisions at \snn = 5.02 \TeV. The statistical and systematic uncertainties are represented by vertical bars and boxes, respectively.}
	\label{fig:qcpvspty}
\end{figure}

The nuclear modification factor ${Q}_\mathrm{CP}$ is calculated from the $\kstar$ and $\phi$ yields normalized to $\langle N_\mathrm{coll} \rangle$ in high multiplicity (central) and low multiplicity (peripheral) collisions, as defined by Equation~\ref{eqn:qcp}. Figure~\ref{fig:qcpvspty} shows the ${Q}_\mathrm{CP}$ of  \kstar(red  circles) and \phim(blue squares) mesons as a function of $\pt$ for \mbox{0--10$\%$ / 40--100$\%$} (top panels) and 10--40$\%$
/ 40--100$\%$ (bottom panels) in various rapidity intervals within the range \mbox{$-$1.2 $<$ $y$ $<$ 0.3} for $\pPb$ collisions at \snn = 5.02 \TeV.
The  ${Q}_\mathrm{CP}$ of $\phim$ mesons seems to be slightly higher than the \kstar one for the ratio of 0--10$\%$ / 40--100$\%$, however, the results for the two meson species are consistent within uncertainties for the ratio   
\mbox{10--40$\%$ / 40--100$\%$} for all measured rapidity intervals.
 An enhancement at intermediate \pt (2.2 $< \pt <$ 5.0 \GeV/$c$), reminiscent of the Cronin effect, is seen 
 for \kstar and \phim mesons in the ${Q}_\mathrm{CP}$.  This enhancement is more pronounced at high negative rapidity, i.e., in the Pb-going direction, and for high multiplicity events. The more pronounced Cronin-like enhancement for the 0-10$\%$ multiplicity class suggests that multiple scattering effects are more relevant for high multiplicity (central) collisions. At high \pt ($>$ 5  \GeV/$c$), the $Q_\mathrm{CP}$ values are greater than unity, which is a known feature of $Q_\mathrm{pPb}$ \footnote[1] {It is defined as the ratio of the yield per equivalent number of nucleon--nucleon collisions at a given centrality or multiplicity class in p--Pb collisions to the yield in minimum-bias pp collisions at the same center-of-mass energy.} and $Q_\mathrm{CP}$ when the centrality or multiplicity classes are defined with the V0 detector, and it is interpreted as a selection bias due to the multiplicity estimator~\cite{ALICE:2014xsp}. The results for \kstar and \phim mesons are consistent between each other within uncertainties. 

\begin{figure}[H]
   	\begin{center}
   	\includegraphics[scale = 0.7]{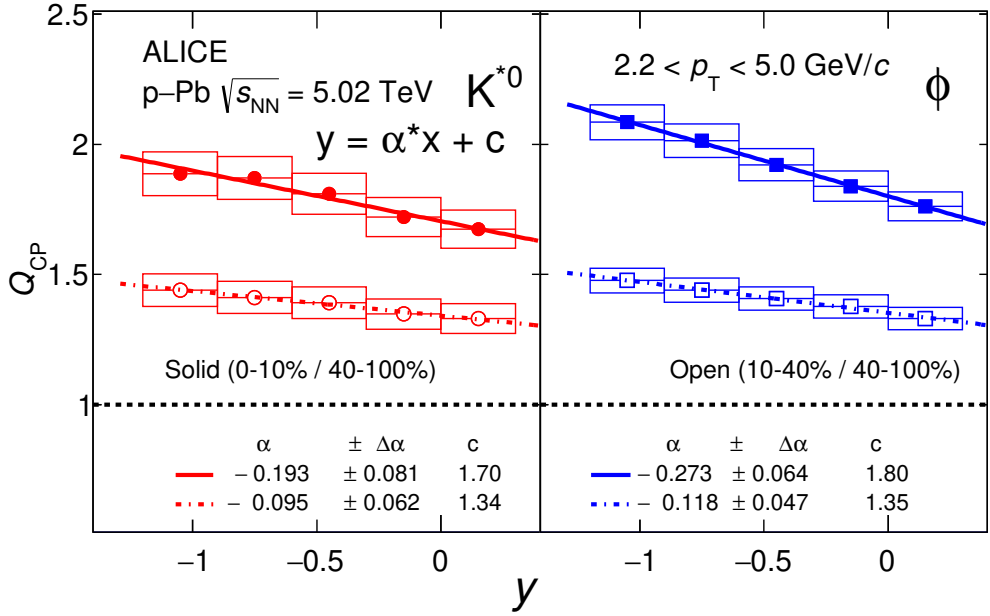}
   	\end{center}
    \caption{The $Q_\mathrm{CP}$ of \kstar (red circles) and \phim(blue squares) mesons as a function of rapidity for \mbox{0--10$\%$ / 40--100$\%$} (solid markers) and 10--40$\%$ / 40--100$\%$ (open markers) in \pPb collisions at \mbox{\snn $=$ 5.02 \TeV}. The solid and doted lines represents the linear fit to data. The statistical and systematic  uncertainties are represented by vertical bars and boxes on the measurements, respectively.}
   	\label{fig:slope}
   \end{figure}

To quantify the rapidity dependence of the nuclear modification factor, the ${Q}_\mathrm{CP}$ values of \kstar and \phim mesons for intermediate $\pt$ (2.2 $< \pt <$ 5.0 \GeV/$c$) are shown as a function of rapidity in Fig.~\ref{fig:slope}. 
The values of ${Q}_\mathrm{CP}$ at intermediate $\pt$ show a faster decrease from the rapidity interval $-$1.2 $< y <$ $-$ 0.9 to  \mbox{0 $< y < $ 0.3}  for  0--10$\%$ / 40--100$\%$  than for 10--40$\%$ / 40--100$\%$, indicating a stronger rapidity dependence of the Cronin-like enhancement in events with high multiplicity.
The stronger rapidity dependence for 0--10$\%$ / 40--100$\%$ can be inferred from the slope parameter ($\alpha$) of the linear function fit to the ${Q}_\mathrm{CP}$ of \kstar and \phim mesons reported in Fig.~\ref{fig:slope}. The slope of the \phim meson ${Q}_\mathrm{CP}$ is slightly larger than the \kstar one. 
A similar conclusion on the $\eta$ dependence of nuclear modification factors of charged hadrons was reported by the BRAHMS Collaboration~\cite{BRAHMS:2004xry}.  

\section{Summary} \label{section:summary}
 The transverse momentum differential yields of $\kstar$ and $\phim$ mesons have been measured in the 
 rapidity interval $-$1.2 $< y <$ 0.3 for various multiplicity classes over the transverse momentum range 
 \mbox{0.8 $< \pt <$16 \GeV/$c$} in $\pPb$ collisions at $\snn$ $=$ 5.02 \TeV with  the ALICE detector.
 The \pt spectra of $\kstar$ and $\phim$ mesons show a multiplicity and rapidity dependence at low \pt, whereas the spectral shapes are similar for all multiplicity classes and rapidity intervals at high \pt($>$ 5 \GeV/${c}$).
 This suggests that nuclear effects influence $\kstar$ and $\phim$ meson production at low \pt.  
 The (d$N/$\rm{d}$y$)/(d$N/$\rm{d}$y)_{y=0}$ ratios decreases with increasing rapidity in the measured interval -1.2 $< y < $0.3, whereas the average transverse momentum $\left(\langle\pt\rangle\right)$ and the $\langle\pt\rangle$/$\langle\pt\rangle_{y=0}$ ratios show a flat behavior for both \kstar and \phim mesons. The rapidity dependence of $\mathrm{d}N/\mathrm{d}y$, $\langle\pt\rangle$ and their ratios with respect to the corresponding values at $y$ $\mathrm{=0}$  are compared with model predictions for minimum-bias events. The EPOS-LHC model, which includes parameterized flow, provides the best description for the magnitudes of $\kstar$ and $\phim$ $\mathrm{d}N/\mathrm{d}y$ and $\langle\pt\rangle$,  whereas  HIJING predictions are in closest agreement with the measured rapidity dependence, which is studied via the ratios (d$N/$\rm{d}$y$)/(d$N/$\rm{d}$y)_{y=0}$ and $\langle\pt\rangle$/$\langle\pt\rangle_{y=0}$.
 The  $Y_\mathrm{asym}$ ratios for \kstar and \phim mesons as a function of \pt show deviations from unity at low \pt for high multiplicity events, while, their values are consistent with unity within uncertainties at high \pt in the measured multiplicity and rapidity intervals. The $Y_\mathrm{asym}$ ratios of \kstar and \phim mesons are found to be consistent between each other within uncertainties in the measured kinematic region. The measured deviations of $Y_\mathrm{asym}$ from unity at low \pt suggest the presence of rapidity dependent nuclear effects such as multiple scattering, nuclear shadowing, parton saturation, and energy loss in cold nuclear matter. None of the models presented here is able to describe the $Y_\mathrm{asym}$ of \kstar and \phim mesons at low \pt.
The nuclear modification factors between the central and peripheral collisions ${Q}_\mathrm{CP}$ for \kstar and \phim mesons as a function of $\pt$ show a bump, with a maximum around \pt=3 \GeV/$c$, suggestive of the Cronin effect. This Cronin-like enhancement is more pronounced for large negative rapidities (in the Pb-going direction) and for more central (higher multiplicity) collisions. The measurements reported in this paper confirm that nuclear effects play an important role in particle production in p--Pb collisions at the LHC energies. They will contribute, along with previous and upcoming measurements of other hadron species, to constrain models and event generators. 

\newenvironment{acknowledgement}{\relax}{\relax}
\begin{acknowledgement}
\section*{Acknowledgements}

The ALICE Collaboration would like to thank all its engineers and technicians for their invaluable contributions to the construction of the experiment and the CERN accelerator teams for the outstanding performance of the LHC complex.
The ALICE Collaboration gratefully acknowledges the resources and support provided by all Grid centres and the Worldwide LHC Computing Grid (WLCG) collaboration.
The ALICE Collaboration acknowledges the following funding agencies for their support in building and running the ALICE detector:
A. I. Alikhanyan National Science Laboratory (Yerevan Physics Institute) Foundation (ANSL), State Committee of Science and World Federation of Scientists (WFS), Armenia;
Austrian Academy of Sciences, Austrian Science Fund (FWF): [M 2467-N36] and Nationalstiftung f\"{u}r Forschung, Technologie und Entwicklung, Austria;
Ministry of Communications and High Technologies, National Nuclear Research Center, Azerbaijan;
Conselho Nacional de Desenvolvimento Cient\'{\i}fico e Tecnol\'{o}gico (CNPq), Financiadora de Estudos e Projetos (Finep), Funda\c{c}\~{a}o de Amparo \`{a} Pesquisa do Estado de S\~{a}o Paulo (FAPESP) and Universidade Federal do Rio Grande do Sul (UFRGS), Brazil;
Bulgarian Ministry of Education and Science, within the National Roadmap for Research Infrastructures 2020-2027 (object CERN), Bulgaria;
Ministry of Education of China (MOEC) , Ministry of Science \& Technology of China (MSTC) and National Natural Science Foundation of China (NSFC), China;
Ministry of Science and Education and Croatian Science Foundation, Croatia;
Centro de Aplicaciones Tecnol\'{o}gicas y Desarrollo Nuclear (CEADEN), Cubaenerg\'{\i}a, Cuba;
Ministry of Education, Youth and Sports of the Czech Republic, Czech Republic;
The Danish Council for Independent Research | Natural Sciences, the VILLUM FONDEN and Danish National Research Foundation (DNRF), Denmark;
Helsinki Institute of Physics (HIP), Finland;
Commissariat \`{a} l'Energie Atomique (CEA) and Institut National de Physique Nucl\'{e}aire et de Physique des Particules (IN2P3) and Centre National de la Recherche Scientifique (CNRS), France;
Bundesministerium f\"{u}r Bildung und Forschung (BMBF) and GSI Helmholtzzentrum f\"{u}r Schwerionenforschung GmbH, Germany;
General Secretariat for Research and Technology, Ministry of Education, Research and Religions, Greece;
National Research, Development and Innovation Office, Hungary;
Department of Atomic Energy Government of India (DAE), Department of Science and Technology, Government of India (DST), University Grants Commission, Government of India (UGC) and Council of Scientific and Industrial Research (CSIR), India;
National Research and Innovation Agency - BRIN, Indonesia;
Istituto Nazionale di Fisica Nucleare (INFN), Italy;
Japanese Ministry of Education, Culture, Sports, Science and Technology (MEXT) and Japan Society for the Promotion of Science (JSPS) KAKENHI, Japan;
Consejo Nacional de Ciencia (CONACYT) y Tecnolog\'{i}a, through Fondo de Cooperaci\'{o}n Internacional en Ciencia y Tecnolog\'{i}a (FONCICYT) and Direcci\'{o}n General de Asuntos del Personal Academico (DGAPA), Mexico;
Nederlandse Organisatie voor Wetenschappelijk Onderzoek (NWO), Netherlands;
The Research Council of Norway, Norway;
Commission on Science and Technology for Sustainable Development in the South (COMSATS), Pakistan;
Pontificia Universidad Cat\'{o}lica del Per\'{u}, Peru;
Ministry of Education and Science, National Science Centre and WUT ID-UB, Poland;
Korea Institute of Science and Technology Information and National Research Foundation of Korea (NRF), Republic of Korea;
Ministry of Education and Scientific Research, Institute of Atomic Physics, Ministry of Research and Innovation and Institute of Atomic Physics and University Politehnica of Bucharest, Romania;
Ministry of Education, Science, Research and Sport of the Slovak Republic, Slovakia;
National Research Foundation of South Africa, South Africa;
Swedish Research Council (VR) and Knut \& Alice Wallenberg Foundation (KAW), Sweden;
European Organization for Nuclear Research, Switzerland;
Suranaree University of Technology (SUT), National Science and Technology Development Agency (NSTDA), Thailand Science Research and Innovation (TSRI) and National Science, Research and Innovation Fund (NSRF), Thailand;
Turkish Energy, Nuclear and Mineral Research Agency (TENMAK), Turkey;
National Academy of  Sciences of Ukraine, Ukraine;
Science and Technology Facilities Council (STFC), United Kingdom;
National Science Foundation of the United States of America (NSF) and United States Department of Energy, Office of Nuclear Physics (DOE NP), United States of America.
In addition, individual groups or members have received support from:
Marie Sk\l{}odowska Curie, Strong 2020 - Horizon 2020, European Research Council (grant nos. 824093, 896850, 950692), European Union;
Academy of Finland (Center of Excellence in Quark Matter) (grant nos. 346327, 346328), Finland;
Programa de Apoyos para la Superaci\'{o}n del Personal Acad\'{e}mico, UNAM, Mexico.

\end{acknowledgement}
\bibliographystyle{utphys}   
\bibliography{RapKstarPhipPb5p02TeV}

\providecommand{\href}[2]{#2}\begingroup\raggedright\begin{thebibliography}{10}

\bibitem{Gyulassy:2004zy}
M.~Gyulassy and L.~McLerran, ``{New forms of QCD matter discovered at RHIC}'',
  \href{http://dx.doi.org/10.1016/j.nuclphysa.2004.10.034}{{\em Nucl. Phys. A}
  {\bfseries 750} (2005) 30--63},
  \href{http://arxiv.org/abs/nucl-th/0405013}{{\ttfamily
  arXiv:nucl-th/0405013}}.

\bibitem{Adams:2005dq}
{\bfseries STAR} Collaboration, J.~Adams {\em et~al.}, ``{Experimental and
  theoretical challenges in the search for the quark gluon plasma: The STAR
  Collaboration's critical assessment of the evidence from RHIC collisions}'',
  \href{http://dx.doi.org/10.1016/j.nuclphysa.2005.03.085}{{\em Nucl. Phys. A}
  {\bfseries 757} (2005) 102--183},
\href{http://arxiv.org/abs/nucl-ex/0501009}{{\ttfamily arXiv:nucl-ex/0501009
  [nucl-ex]}}.

\bibitem{Schukraft:2011na}
{\bfseries ALICE} Collaboration, J.~Schukraft, ``{Heavy Ion physics with the
  ALICE experiment at the CERN LHC}'',
  \href{http://dx.doi.org/10.1098/rsta.2011.0469}{{\em Phil. Trans. Roy. Soc.
  Lond. A} {\bfseries 370} (2012) 917--932},
  \href{http://arxiv.org/abs/1109.4291}{{\ttfamily arXiv:1109.4291 [hep-ex]}}.

\bibitem{Braun-Munzinger:2015hba}
P.~Braun-Munzinger, V.~Koch, T.~Sch\"afer, and J.~Stachel, ``{Properties of hot
  and dense matter from relativistic heavy ion collisions}'',
  \href{http://dx.doi.org/10.1016/j.physrep.2015.12.003}{{\em Phys. Rept.}
  {\bfseries 621} (2016) 76--126},
  \href{http://arxiv.org/abs/1510.00442}{{\ttfamily arXiv:1510.00442
  [nucl-th]}}.

\bibitem{PHENIX:2003qdw}
{\bfseries PHENIX} Collaboration, S.~S. Adler {\em et~al.}, ``{Absence of
  suppression in particle production at large transverse momentum in
  $\sqrt{s_\mathrm{NN}}$ = 200 GeV d + Au collisions}'',
  \href{http://dx.doi.org/10.1103/PhysRevLett.91.072303}{{\em Phys. Rev. Lett.}
  {\bfseries 91} (2003) 072303},
  \href{http://arxiv.org/abs/nucl-ex/0306021}{{\ttfamily
  arXiv:nucl-ex/0306021}}.

\bibitem{PHOBOS:2003uzz}
{\bfseries PHOBOS} Collaboration, B.~B. Back {\em et~al.}, ``{Centrality
  dependence of charged hadron transverse momentum spectra in d + Au collisions
  at $\sqrt{s_\mathrm{NN}}$ = 200 GeV}'',
  \href{http://dx.doi.org/10.1103/PhysRevLett.91.072302}{{\em Phys. Rev. Lett.}
  {\bfseries 91} (2003) 072302},
  \href{http://arxiv.org/abs/nucl-ex/0306025}{{\ttfamily
  arXiv:nucl-ex/0306025}}.

\bibitem{BRAHMS:2004xry}
{\bfseries BRAHMS} Collaboration, I.~Arsene {\em et~al.}, ``{On the evolution
  of the nuclear modification factors with rapidity and centrality in d + Au
  collisions at $\sqrt{s_\mathrm{NN}}$ = 200 GeV}'',
  \href{http://dx.doi.org/10.1103/PhysRevLett.93.242303}{{\em Phys. Rev. Lett.}
  {\bfseries 93} (2004) 242303},
  \href{http://arxiv.org/abs/nucl-ex/0403005}{{\ttfamily
  arXiv:nucl-ex/0403005}}.

\bibitem{ZEUS:1997etp}
{\bfseries ZEUS} Collaboration, J.~Breitweg {\em et~al.}, ``{Measurement of the
  proton structure function F$_{2}$ and $\sigma_{tot}$($\gamma^*$p) at low
  $q^2$ and very low x at HERA}'',
  \href{http://dx.doi.org/10.1016/S0370-2693(97)00905-2}{{\em Phys. Lett. B}
  {\bfseries 407} (1997) 432--448},
  \href{http://arxiv.org/abs/hep-ex/9707025}{{\ttfamily arXiv:hep-ex/9707025}}.

\bibitem{ALICE:2012mj}
{\bfseries ALICE} Collaboration, B.~Abelev {\em et~al.}, ``{Transverse momentum
  distribution and nuclear modification factor of charged particles in p--Pb
  collisions at $\sqrt{s_{\mathrm{NN}}}=5.02$ TeV}'',
  \href{http://dx.doi.org/10.1103/PhysRevLett.110.082302}{{\em Phys. Rev.
  Lett.} {\bfseries 110} (2013) 082302},
  \href{http://arxiv.org/abs/1210.4520}{{\ttfamily arXiv:1210.4520 [nucl-ex]}}.

\bibitem{Acharya:2018qsh}
{\bfseries ALICE} Collaboration, S.~Acharya {\em et~al.}, ``{Transverse
  momentum spectra and nuclear modification factors of charged particles in pp,
  p-Pb and Pb-Pb collisions at the LHC}'',
  \href{http://dx.doi.org/10.1007/JHEP11(2018)013}{{\em JHEP} {\bfseries 11}
  (2018) 013},
\href{http://arxiv.org/abs/1802.09145}{{\ttfamily arXiv:1802.09145 [nucl-ex]}}.

\bibitem{Adams:2004ep}
{\bfseries STAR} Collaboration, J.~Adams {\em et~al.}, ``{K$^{*}(892)^{0}$
  resonance production in Au+Au and p+p collisions at
  $\sqrt{s_\mathrm{NN}}=200$ GeV at STAR}'',
  \href{http://dx.doi.org/10.1103/PhysRevC.71.064902}{{\em Phys. Rev. C}
  {\bfseries 71} (2005) 064902},
\href{http://arxiv.org/abs/nucl-ex/0412019}{{\ttfamily arXiv:nucl-ex/0412019
  [nucl-ex]}}.

\bibitem{Anticic:2011zr}
{\bfseries NA49} Collaboration, T.~Anticic {\em et~al.}, ``{$K^{\ast}(892)^0$
  and $\bar{K}^{\ast}(892)^0$ production in central Pb+Pb, Si+Si, C+C and
  inelastic p+p collisions at 158$A$~GeV}'',
  \href{http://dx.doi.org/10.1103/PhysRevC.84.064909}{{\em Phys. Rev. C}
  {\bfseries 84} (2011) 064909},
\href{http://arxiv.org/abs/1105.3109}{{\ttfamily arXiv:1105.3109 [nucl-ex]}}.

\bibitem{Acharya:2019yoi}
{\bfseries ALICE} Collaboration, S.~Acharya {\em et~al.}, ``{Production of
  charged pions, kaons, and (anti-)protons in Pb-Pb and inelastic $pp$
  collisions at $\sqrt {s_\mathrm{NN}}$ = 5.02 TeV}'',
  \href{http://dx.doi.org/10.1103/PhysRevC.101.044907}{{\em Phys. Rev. C}
  {\bfseries 101} (2020) 044907},
\href{http://arxiv.org/abs/1910.07678}{{\ttfamily arXiv:1910.07678 [nucl-ex]}}.

\bibitem{ALICE:2021ptz}
{\bfseries ALICE} Collaboration, S.~Acharya {\em et~al.}, ``{Production of
  K$^{*}(892)^{0}$ and $\phi(1020)$ in pp and Pb-Pb collisions at
  $\sqrt{s_{\mathrm{NN}}} = 5.02$ TeV}'',
  \href{http://dx.doi.org/10.1103/PhysRevC.106.034907}{{\em Phys. Rev. C}
  {\bfseries 106} (2022) 034907},
  \href{http://arxiv.org/abs/2106.13113}{{\ttfamily arXiv:2106.13113
  [nucl-ex]}}.

\bibitem{Albacete:2016veq}
J.~L. Albacete {\em et~al.}, ``{Predictions for $p+$Pb Collisions at
  $\sqrt{s_\mathrm{NN}} = 5$ TeV: Comparison with Data}'',
  \href{http://dx.doi.org/10.1142/S0218301316300058}{{\em Int. J. Mod. Phys. E}
  {\bfseries 25} (2016) 1630005},
  \href{http://arxiv.org/abs/1605.09479}{{\ttfamily arXiv:1605.09479
  [hep-ph]}}.

\bibitem{CMS:2016zzh}
{\bfseries CMS} Collaboration, V.~Khachatryan {\em et~al.}, ``{Multiplicity and
  rapidity dependence of strange hadron production in pp, pPb, and PbPb
  collisions at the LHC}'',
  \href{http://dx.doi.org/10.1016/j.physletb.2017.01.075}{{\em Phys. Lett. B}
  {\bfseries 768} (2017) 103--129},
  \href{http://arxiv.org/abs/1605.06699}{{\ttfamily arXiv:1605.06699
  [nucl-ex]}}.

\bibitem{CMS:2019isl}
{\bfseries CMS} Collaboration, A.~M. Sirunyan {\em et~al.}, ``{Strange hadron
  production in pp and pPb collisions at $\sqrt{s_\mathrm{NN}}= $ 5.02 TeV}'',
  \href{http://dx.doi.org/10.1103/PhysRevC.101.064906}{{\em Phys. Rev. C}
  {\bfseries 101} (2020) 064906},
  \href{http://arxiv.org/abs/1910.04812}{{\ttfamily arXiv:1910.04812
  [hep-ex]}}.

\bibitem{STAR:2006kxj}
{\bfseries STAR} Collaboration, B.~I. Abelev {\em et~al.}, ``{Rapidity and
  species dependence of particle production at large transverse momentum for
  d+Au collisions at $\sqrt{s_{\mathrm{NN }}}$ = 200 GeV}'',
  \href{http://dx.doi.org/10.1103/PhysRevC.76.054903}{{\em Phys. Rev. C}
  {\bfseries 76} (2007) 054903},
  \href{http://arxiv.org/abs/nucl-ex/0609021}{{\ttfamily
  arXiv:nucl-ex/0609021}}.

\bibitem{PHENIX:2018hho}
{\bfseries PHENIX} Collaboration, A.~Adare {\em et~al.}, ``{Pseudorapidity
  Dependence of Particle Production and Elliptic Flow in Asymmetric Nuclear
  Collisions of $p+$Al, $p+$Au, $d+$Au, and $^{3}$He$+$Au at
  $\sqrt{s_{_{NN}}}=200$ GeV}'',
  \href{http://dx.doi.org/10.1103/PhysRevLett.121.222301}{{\em Phys. Rev.
  Lett.} {\bfseries 121} (2018) 222301},
  \href{http://arxiv.org/abs/1807.11928}{{\ttfamily arXiv:1807.11928
  [nucl-ex]}}.

\bibitem{PHENIX:2014fnc}
{\bfseries PHENIX} Collaboration, A.~Adare {\em et~al.}, ``{Measurement of
  long-range angular correlation and quadrupole anisotropy of pions and
  (anti)protons in central $d+$Au collisions at $\sqrt{s_\mathrm{NN}}$=200
  GeV}'', \href{http://dx.doi.org/10.1103/PhysRevLett.114.192301}{{\em Phys.
  Rev. Lett.} {\bfseries 114} (2015) 192301},
  \href{http://arxiv.org/abs/1404.7461}{{\ttfamily arXiv:1404.7461 [nucl-ex]}}.

\bibitem{PHENIX:2016cfs}
{\bfseries PHENIX} Collaboration, C.~Aidala {\em et~al.}, ``{Measurement of
  long-range angular correlations and azimuthal anisotropies in
  high-multiplicity $p+$Au collisions at $\sqrt{s_\mathrm{NN}}=200$ GeV}'',
  \href{http://dx.doi.org/10.1103/PhysRevC.95.034910}{{\em Phys. Rev. C}
  {\bfseries 95} (2017) 034910},
  \href{http://arxiv.org/abs/1609.02894}{{\ttfamily arXiv:1609.02894
  [nucl-ex]}}.

\bibitem{Kovchegov:2004jm}
{\relax Yu}.~V. Kovchegov, ``{Cronin effect and high-p$_{T}$ suppression in
  p(d)A collisions}'',
\href{http://dx.doi.org/10.1088/0954-3899/30/8/042}{{\em J. Phys. G} {\bfseries
  30} (2004) S979--S982}.

\bibitem{ALICE:2016fzo}
{\bfseries ALICE} Collaboration, J.~Adam {\em et~al.}, ``{Enhanced production
  of multi-strange hadrons in high-multiplicity proton-proton collisions}'',
  \href{http://dx.doi.org/10.1038/nphys4111}{{\em Nature Phys.} {\bfseries 13}
  (2017) 535--539}, \href{http://arxiv.org/abs/1606.07424}{{\ttfamily
  arXiv:1606.07424 [nucl-ex]}}.

\bibitem{ALICE:2012eyl}
{\bfseries ALICE} Collaboration, B.~Abelev {\em et~al.}, ``{Long-range angular
  correlations on the near and away side in $p$-Pb collisions at
  $\sqrt{s_\mathrm{NN}}=5.02$ TeV}'',
  \href{http://dx.doi.org/10.1016/j.physletb.2013.01.012}{{\em Phys. Lett. B}
  {\bfseries 719} (2013) 29--41},
  \href{http://arxiv.org/abs/1212.2001}{{\ttfamily arXiv:1212.2001 [nucl-ex]}}.

\bibitem{PHENIX:2018lia}
{\bfseries PHENIX} Collaboration, C.~Aidala {\em et~al.}, ``{Creation of
  quark\textendash{}gluon plasma droplets with three distinct geometries}'',
  \href{http://dx.doi.org/10.1038/s41567-018-0360-0}{{\em Nature Phys.}
  {\bfseries 15} (2019) 214--220},
  \href{http://arxiv.org/abs/1805.02973}{{\ttfamily arXiv:1805.02973
  [nucl-ex]}}.

\bibitem{Adam:2016bpr}
{\bfseries ALICE} Collaboration, J.~Adam {\em et~al.}, ``{Production of K$^{*}$
  (892)$^{0}$ and $\phi $ (1020) in p--Pb collisions at
  $\sqrt{s_{{\mathrm{NN}}}}$ = 5.02 TeV}'',
  \href{http://dx.doi.org/10.1140/epjc/s10052-016-4088-7}{{\em Eur. Phys. J. C}
  {\bfseries 76} (2016) 245},
\href{http://arxiv.org/abs/1601.07868}{{\ttfamily arXiv:1601.07868 [nucl-ex]}}.

\bibitem{ALICE:2021uyz}
{\bfseries ALICE} Collaboration, S.~Acharya {\em et~al.},
  ``{$\mathrm{K}^{*}(\mathrm{892})^{0}$ and $\mathrm{\phi(1020)}$ production in
  p-Pb collisions at $\sqrt{s_{\mathrm{NN}}}$ = 8.16 TeV}'',
  \href{http://dx.doi.org/10.1103/PhysRevC.107.055201}{{\em Phys. Rev. C}
  {\bfseries 107} (2023) 055201},
  \href{http://arxiv.org/abs/2110.10042}{{\ttfamily arXiv:2110.10042
  [nucl-ex]}}.

\bibitem{ALICE:2021ucq}
{\bfseries ALICE} Collaboration, S.~Acharya {\em et~al.}, ``{Energy dependence
  of $\phi $ meson production at forward rapidity in pp collisions at the
  LHC}'', \href{http://dx.doi.org/10.1140/epjc/s10052-021-09545-3}{{\em Eur.
  Phys. J. C} {\bfseries 81} (2021) 772},
  \href{http://arxiv.org/abs/2105.00713}{{\ttfamily arXiv:2105.00713
  [nucl-ex]}}.

\bibitem{ALICE:2015cql}
{\bfseries ALICE} Collaboration, J.~Adam {\em et~al.}, ``{$\phi$-meson
  production at forward rapidity in p-Pb collisions at $\sqrt{s_{\mathrm{NN}}}$
  = 5.02 TeV and in pp collisions at $\sqrt{s}$ = 2.76 TeV}'',
  \href{http://dx.doi.org/10.1016/j.physletb.2017.01.074}{{\em Phys. Lett. B}
  {\bfseries 768} (2017) 203--217},
  \href{http://arxiv.org/abs/1506.09206}{{\ttfamily arXiv:1506.09206
  [nucl-ex]}}.

\bibitem{Kang:2012kc}
Z.-B. Kang, I.~Vitev, and H.~Xing, ``{Nuclear modification of high transverse
  momentum particle production in p+A collisions at RHIC and LHC}'',
  \href{http://dx.doi.org/10.1016/j.physletb.2012.10.046}{{\em Phys. Lett. B}
  {\bfseries 718} (2012) 482--487},
\href{http://arxiv.org/abs/1209.6030}{{\ttfamily arXiv:1209.6030 [hep-ph]}}.

\bibitem{ALICE:2013wgn}
{\bfseries ALICE} Collaboration, B.~B. Abelev {\em et~al.}, ``{Multiplicity
  Dependence of Pion, Kaon, Proton and Lambda Production in p-Pb Collisions at
  $\sqrt{s_\mathrm{NN}}$ = 5.02 TeV}'',
  \href{http://dx.doi.org/10.1016/j.physletb.2013.11.020}{{\em Phys. Lett. B}
  {\bfseries 728} (2014) 25--38},
  \href{http://arxiv.org/abs/1307.6796}{{\ttfamily arXiv:1307.6796 [nucl-ex]}}.

\bibitem{Bozek:2015swa}
P.~Bozek, A.~Bzdak, and G.-L. Ma, ``{Rapidity dependence of elliptic and
  triangular flow in proton\textendash{}nucleus collisions from collective
  dynamics}'', \href{http://dx.doi.org/10.1016/j.physletb.2015.06.007}{{\em
  Phys. Lett. B} {\bfseries 748} (2015) 301--305},
  \href{http://arxiv.org/abs/1503.03655}{{\ttfamily arXiv:1503.03655
  [hep-ph]}}.

\bibitem{Bozek:2013sda}
P.~Bozek, A.~Bzdak, and V.~Skokov, ``{The rapidity dependence of the average
  transverse momentum in p+Pb collisions at the LHC: the Color Glass Condensate
  versus hydrodynamics}'',
  \href{http://dx.doi.org/10.1016/j.physletb.2013.12.034}{{\em Phys. Lett. B}
  {\bfseries 728} (2014) 662--665},
  \href{http://arxiv.org/abs/1309.7358}{{\ttfamily arXiv:1309.7358 [hep-ph]}}.

\bibitem{Acharya:2019bli}
{\bfseries ALICE} Collaboration, S.~Acharya {\em et~al.}, ``{Multiplicity
  dependence of K*(892)$^{0}$ and $\phi$(1020) production in pp collisions at
  $\sqrt {s}$ = 13 TeV}'',
  \href{http://dx.doi.org/10.1016/j.physletb.2020.135501}{{\em Phys. Lett. B}
  {\bfseries 807} (2020) 135501},
\href{http://arxiv.org/abs/1910.14397}{{\ttfamily arXiv:1910.14397 [nucl-ex]}}.

\bibitem{Pierog:2013ria}
T.~Pierog, I.~Karpenko, J.~M. Katzy, E.~Yatsenko, and K.~Werner, ``{EPOS LHC:
  Test of collective hadronization with data measured at the CERN Large Hadron
  Collider}'', \href{http://dx.doi.org/10.1103/PhysRevC.92.034906}{{\em Phys.
  Rev. C} {\bfseries 92} (2015) 034906},
\href{http://arxiv.org/abs/1306.0121}{{\ttfamily arXiv:1306.0121 [hep-ph]}}.

\bibitem{Werner:2013yia}
K.~Werner, L.~Karpenko, M.~Bleicher, and T.~Pierog, ``{The Physics of EPOS}'',
\href{http://dx.doi.org/10.1051/epjconf/20125205001}{{\em EPJ Web Conf.}
  {\bfseries 52} (2013) 05001}.

\bibitem{Pierog:2009zt}
T.~Pierog and K.~Werner, ``{EPOS Model and Ultra High Energy Cosmic Rays}'',
  \href{http://dx.doi.org/10.1016/j.nuclphysbps.2009.09.017}{{\em Nucl. Phys.
  Proc. Suppl.} {\bfseries 196} (2009) 102--105},
\href{http://arxiv.org/abs/0905.1198}{{\ttfamily arXiv:0905.1198 [hep-ph]}}.

\bibitem{Werner:2013tya}
K.~Werner, B.~Guiot, I.~Karpenko, and T.~Pierog, ``{Analysing radial flow
  features in p-Pb and p-p collisions at several TeV by studying identified
  particle production in EPOS3}'',
  \href{http://dx.doi.org/10.1103/PhysRevC.89.064903}{{\em Phys. Rev. C}
  {\bfseries 89} (2014) 064903},
\href{http://arxiv.org/abs/1312.1233}{{\ttfamily arXiv:1312.1233 [nucl-th]}}.

\bibitem{Knospe:2021jgt}
A.~G. Knospe, C.~Markert, K.~Werner, J.~Steinheimer, and M.~Bleicher,
  ``{Hadronic resonance production and interaction in p-Pb collisions at LHC
  energies in EPOS3}'',
  \href{http://dx.doi.org/10.1103/PhysRevC.104.054907}{{\em Phys. Rev. C}
  {\bfseries 104} (2021) 054907},
  \href{http://arxiv.org/abs/2102.06797}{{\ttfamily arXiv:2102.06797
  [nucl-th]}}.

\bibitem{Roesler:2000he}
S.~Roesler, R.~Engel, and J.~Ranft,
  \href{http://dx.doi.org/10.1007/978-3-642-18211-2_166}{``{The Monte Carlo
  event generator DPMJET-III}'',}
\newblock 12, 2000.
\newblock \href{http://arxiv.org/abs/hep-ph/0012252}{{\ttfamily
  arXiv:hep-ph/0012252}}.

\bibitem{Gyulassy:1994ew}
M.~Gyulassy and X.-N. Wang, ``{HIJING 1.0: A Monte Carlo program for parton and
  particle production in high-energy hadronic and nuclear collisions}'',
  \href{http://dx.doi.org/10.1016/0010-4655(94)90057-4}{{\em Comput. Phys.
  Commun.} {\bfseries 83} (1994) 307},
\href{http://arxiv.org/abs/nucl-th/9502021}{{\ttfamily arXiv:nucl-th/9502021
  [nucl-th]}}.

\bibitem{Bierlich:2018xfw}
C.~Bierlich, G.~Gustafson, L.~L\"onnblad, and H.~Shah, ``{The Angantyr model
  for Heavy-Ion Collisions in PYTHIA8}'',
  \href{http://dx.doi.org/10.1007/JHEP10(2018)134}{{\em JHEP} {\bfseries 10}
  (2018) 134}, \href{http://arxiv.org/abs/1806.10820}{{\ttfamily
  arXiv:1806.10820 [hep-ph]}}.

\bibitem{Pi:1992ug}
H.~Pi, ``{An Event generator for interactions between hadrons and nuclei:
  FRITIOF version 7.0}'',
  \href{http://dx.doi.org/10.1016/0010-4655(92)90082-A}{{\em Comput. Phys.
  Commun.} {\bfseries 71} (1992) 173--192}.

\bibitem{Aamodt:2008zz}
{\bfseries ALICE} Collaboration, K.~Aamodt {\em et~al.}, ``{The ALICE
  experiment at the CERN LHC}'',
\href{http://dx.doi.org/10.1088/1748-0221/3/08/S08002}{{\em JINST} {\bfseries
  3} (2008) S08002}.

\bibitem{Abelev:2014ffa}
{\bfseries ALICE} Collaboration, B.~Abelev {\em et~al.}, ``{Performance of the
  ALICE Experiment at the CERN LHC}'',
  \href{http://dx.doi.org/10.1142/S0217751X14300440}{{\em Int. J. Mod. Phys. A}
  {\bfseries 29} (2014) 1430044},
\href{http://arxiv.org/abs/1402.4476}{{\ttfamily arXiv:1402.4476 [nucl-ex]}}.

\bibitem{Aamodt:2010aa}
{\bfseries ALICE} Collaboration, K.~Aamodt {\em et~al.}, ``{Alignment of the
  ALICE Inner Tracking System with cosmic-ray tracks}'',
  \href{http://dx.doi.org/10.1088/1748-0221/5/03/P03003}{{\em JINST} {\bfseries
  5} (2010) P03003},
\href{http://arxiv.org/abs/1001.0502}{{\ttfamily arXiv:1001.0502
  [physics.ins-det]}}.

\bibitem{Alme:2010ke}
J.~Alme {\em et~al.}, ``{The ALICE TPC, a large 3-dimensional tracking device
  with fast readout for ultra-high multiplicity events}'',
  \href{http://dx.doi.org/10.1016/j.nima.2010.04.042}{{\em Nucl. Instrum. Meth.
  A} {\bfseries 622} (2010) 316--367},
\href{http://arxiv.org/abs/1001.1950}{{\ttfamily arXiv:1001.1950
  [physics.ins-det]}}.

\bibitem{DeGruttola:2014eti}
{\bfseries ALICE} Collaboration, D.~De~Gruttola, ``{Particle IDentification
  with the ALICE Time-Of-Flight detector at the LHC}'',
  \href{http://dx.doi.org/10.1088/1748-0221/9/10/C10019}{{\em JINST} {\bfseries
  9} (2014) C10019}.

\bibitem{Abbas:2013taa}
{\bfseries ALICE} Collaboration, E.~Abbas {\em et~al.}, ``{Performance of the
  ALICE VZERO system}'',
  \href{http://dx.doi.org/10.1088/1748-0221/8/10/P10016}{{\em JINST} {\bfseries
  8} (2013) P10016},
\href{http://arxiv.org/abs/1306.3130}{{\ttfamily arXiv:1306.3130 [nucl-ex]}}.

\bibitem{Miller:2007ri}
M.~L. Miller, K.~Reygers, S.~J. Sanders, and P.~Steinberg, ``{Glauber modeling
  in high energy nuclear collisions}'',
  \href{http://dx.doi.org/10.1146/annurev.nucl.57.090506.123020}{{\em Ann. Rev.
  Nucl. Part. Sci.} {\bfseries 57} (2007) 205--243},
\href{http://arxiv.org/abs/nucl-ex/0701025}{{\ttfamily arXiv:nucl-ex/0701025
  [nucl-ex]}}.

\bibitem{ALICE:2014xsp}
{\bfseries ALICE} Collaboration, J.~Adam {\em et~al.}, ``{Centrality dependence
  of particle production in p-Pb collisions at $\sqrt{s_{\mathrm{NN}} }$= 5.02
  TeV}'', \href{http://dx.doi.org/10.1103/PhysRevC.91.064905}{{\em Phys. Rev.
  C} {\bfseries 91} (2015) 064905},
  \href{http://arxiv.org/abs/1412.6828}{{\ttfamily arXiv:1412.6828 [nucl-ex]}}.

\bibitem{Tanabashi:2018oca}
{\bfseries Particle Data Group} Collaboration, M.~Tanabashi {\em et~al.},
  ``{Review of Particle Physics}'',
\href{http://dx.doi.org/10.1103/PhysRevD.98.030001}{{\em Phys. Rev. D}
  {\bfseries 98} (2018) 030001}.

\bibitem{Acharya:2019wyb}
{\bfseries ALICE} Collaboration, S.~Acharya {\em et~al.},
  ``{$\rm{K}^{*}(\rm{892})^{0}$ and $\phi(1020)$ production at midrapidity in
  pp collisions at $\sqrt{s}$ = 8 TeV}'',
  \href{http://dx.doi.org/10.1103/PhysRevC.102.024912}{{\em Phys. Rev. C}
  {\bfseries 102} (2020) 024912},
\href{http://arxiv.org/abs/1910.14410}{{\ttfamily arXiv:1910.14410 [nucl-ex]}}.

\bibitem{Brun:1119728}
R.~Brun, F.~Bruyant, M.~Maire, A.~C. McPherson, and P.~Zanarini, {\em {GEANT 3:
  user's guide Geant 3.10, Geant 3.11; rev. version}}.
\newblock CERN, Geneva, 1987.
\newblock \url{https://cds.cern.ch/record/1119728}.

\bibitem{Tsallis:1987eu}
C.~Tsallis, ``{Possible Generalization of Boltzmann-Gibbs Statistics}'',
\href{http://dx.doi.org/10.1007/BF01016429}{{\em J. Statist. Phys.} {\bfseries
  52} (1988) 479--487}.

\bibitem{ALICE:2020jsh}
{\bfseries ALICE} Collaboration, S.~Acharya {\em et~al.}, ``{Production of
  light-flavor hadrons in pp collisions at $\sqrt{s}~=~7\text { and }\sqrt{s} =
  13 \, \text { TeV} $}'',
  \href{http://dx.doi.org/10.1140/epjc/s10052-020-08690-5}{{\em Eur. Phys. J.
  C} {\bfseries 81} (2021) 256},
  \href{http://arxiv.org/abs/2005.11120}{{\ttfamily arXiv:2005.11120
  [nucl-ex]}}.

\end{thebibliography}\endgroup

\newpage
\appendix

\section{Multiplicity and rapidity dependence of d$N/\mathrm{d}y$ and $\langle{p}_{\mathrm{T}}\rangle$} \label{section:appendix} 
\begin{figure}[b]
	\begin{center}
	\includegraphics[scale=0.7]{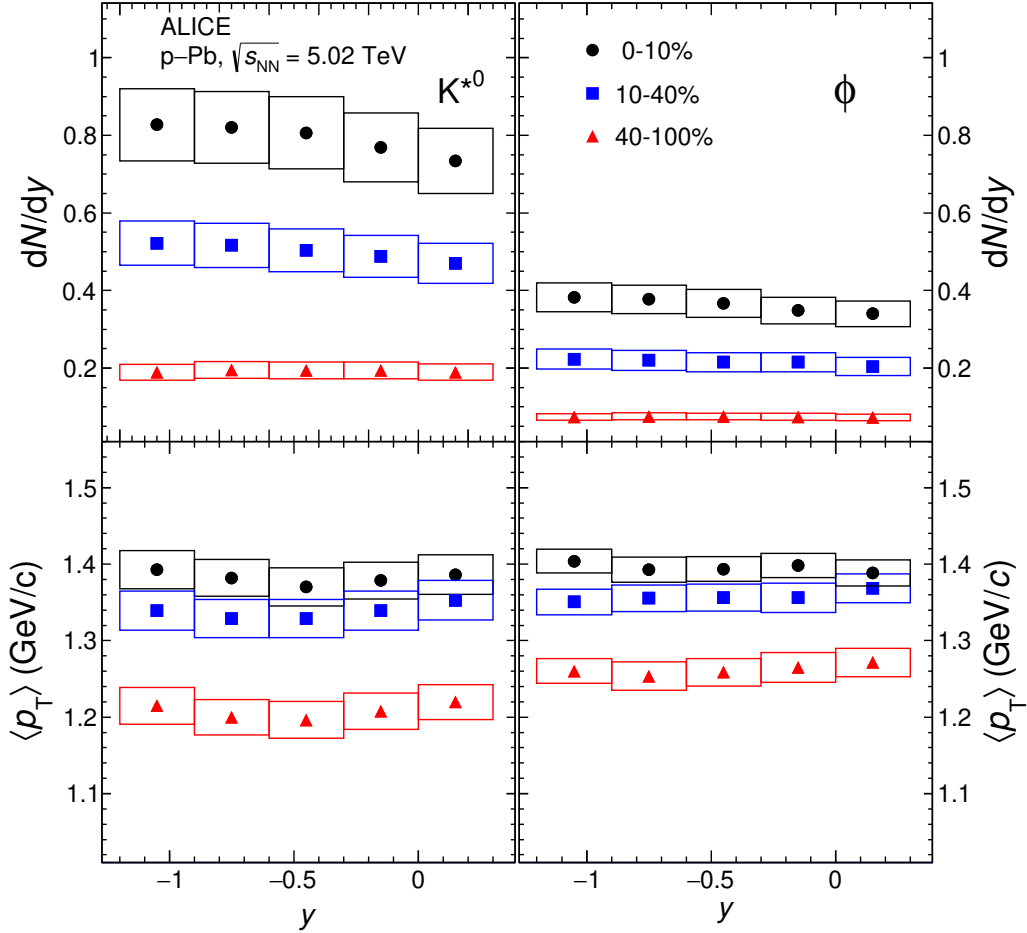}          
	\end{center}
	\caption{The \pt integrated yield (d$N/\mathrm{d}y$) (top panels) and mean transverse momentum  $\left(\langle\pt\rangle\right)$ (bottom panels) for \kstar (left panels) and \phim (right panels) mesons as a function of $y$ measured  for the multiplicity classes 0--10$\%$, 10--40$\%$ and 40--100$\%$ in \pPb collisions at \snn $=$ 5.02 \TeV. The statistical uncertainties are represented as bars whereas boxes indicate the total systematic uncertainties on the measurements. }
	\label{fig:dndymeanptmult}
\end{figure}

Figure~\ref{fig:dndymeanptmult} shows the multiplicity dependence of the d$N/\mathrm{d}y$ and $\langle{p}_{\mathrm{T}}\rangle$  of \kstar and \phim mesons as a function of $y$ in \pPb collisions at $\sqrt{{s}_\mathrm{NN}}$ = 5.02 \TeV.  The d$N/\mathrm{d}y$ and the  $\langle\pt\rangle$ increase with multiplicity for a given rapidity interval. The d$N/\mathrm{d}y$ shows a weak rapidity dependence with large uncertainties, and suggesting a more pronounced dependence for events in the highest multiplicity class (0--10$\%$).
The  $\langle\pt\rangle$ shows a flat behavior as a function of rapidity for all multiplicity classes in the measured rapidity interval. 
Similar behavior in the average transverse kinetic energy as a function of rapidity for strange hadrons was reported in Ref.~\cite{CMS:2016zzh}.

%
%

\newpage
\section{The ALICE Collaboration}
\label{app:collab}
\begin{flushleft} 
\small

S.~Acharya\,\orcidlink{0000-0002-9213-5329}\,$^{\rm 124,131}$, 
D.~Adamov\'{a}\,\orcidlink{0000-0002-0504-7428}\,$^{\rm 86}$, 
A.~Adler$^{\rm 69}$, 
G.~Aglieri Rinella\,\orcidlink{0000-0002-9611-3696}\,$^{\rm 32}$, 
M.~Agnello\,\orcidlink{0000-0002-0760-5075}\,$^{\rm 29}$, 
N.~Agrawal\,\orcidlink{0000-0003-0348-9836}\,$^{\rm 50}$, 
Z.~Ahammed\,\orcidlink{0000-0001-5241-7412}\,$^{\rm 131}$, 
S.~Ahmad\,\orcidlink{0000-0003-0497-5705}\,$^{\rm 15}$, 
S.U.~Ahn\,\orcidlink{0000-0001-8847-489X}\,$^{\rm 70}$, 
I.~Ahuja\,\orcidlink{0000-0002-4417-1392}\,$^{\rm 37}$
% ...

\section*{Affiliation Notes}

$^{\rm I}$ Deceased\\
$^{\rm II}$ Also at: Italian National Agency for New Technologies, Energy and Sustainable Economic Development (ENEA), Bologna, Italy\\
$^{\rm III}$ Also at: Dipartimento DET del Politecnico di Torino, Turin, Italy\\
$^{\rm IV}$ Also at: Department of Applied Physics, Aligarh Muslim University, Aligarh, India\\
$^{\rm V}$ Also at: Institute of Theoretical Physics, University of Wroclaw, Poland\\
$^{\rm VI}$ Also at: University of Kansas, Lawrence, Kansas, United States\\
$^{\rm VII}$ Also at: An institution covered by a cooperation agreement with CERN\\

\section*{Collaboration Institutes}

$^{1}$ A.I. Alikhanyan National Science Laboratory (Yerevan Physics Institute) Foundation, Yerevan, Armenia\\
$^{2}$ AGH University of Science and Technology, Cracow, Poland\\
$^{3}$ Bogolyubov Institute for Theoretical Physics, National Academy of Sciences of Ukraine, Kiev, Ukraine\\
$^{4}$ Bose Institute, Department of Physics  and Centre for Astroparticle Physics and Space Science (CAPSS), Kolkata, India\\
$^{5}$ California Polytechnic State University, San Luis Obispo, California, United States\\
$^{6}$ Central China Normal University, Wuhan, China\\
$^{7}$ Centro de Aplicaciones Tecnol\'{o}gicas y Desarrollo Nuclear (CEADEN), Havana, Cuba\\
$^{8}$ Centro de Investigaci\'{o}n y de Estudios Avanzados (CINVESTAV), Mexico City and M\'{e}rida, Mexico\\
$^{9}$ Chicago State University, Chicago, Illinois, United States\\
$^{10}$ China Institute of Atomic Energy, Beijing, China\\
$^{11}$ Chungbuk National University, Cheongju, Republic of Korea\\
$^{12}$ Comenius University Bratislava, Faculty of Mathematics, Physics and Informatics, Bratislava, Slovak Republic\\
$^{13}$ COMSATS University Islamabad, Islamabad, Pakistan\\
$^{14}$ Creighton University, Omaha, Nebraska, United States\\
$^{15}$ Department of Physics, Aligarh Muslim University, Aligarh, India\\
$^{16}$ Department of Physics, Pusan National University, Pusan, Republic of Korea\\
$^{17}$ Department of Physics, Sejong University, Seoul, Republic of Korea\\
$^{18}$ Department of Physics, University of California, Berkeley, California, United States\\
$^{19}$ Department of Physics, University of Oslo, Oslo, Norway\\
$^{20}$ Department of Physics and Technology, University of Bergen, Bergen, Norway\\
$^{21}$ Dipartimento di Fisica, Universit\`{a} di Pavia, Pavia, Italy\\
$^{22}$ Dipartimento di Fisica dell'Universit\`{a} and Sezione INFN, Cagliari, Italy\\
$^{23}$ Dipartimento di Fisica dell'Universit\`{a} and Sezione INFN, Trieste, Italy\\
$^{24}$ Dipartimento di Fisica dell'Universit\`{a} and Sezione INFN, Turin, Italy\\
$^{25}$ Dipartimento di Fisica e Astronomia dell'Universit\`{a} and Sezione INFN, Bologna, Italy\\
$^{26}$ Dipartimento di Fisica e Astronomia dell'Universit\`{a} and Sezione INFN, Catania, Italy\\
$^{27}$ Dipartimento di Fisica e Astronomia dell'Universit\`{a} and Sezione INFN, Padova, Italy\\
$^{28}$ Dipartimento di Fisica `E.R.~Caianiello' dell'Universit\`{a} and Gruppo Collegato INFN, Salerno, Italy\\
$^{29}$ Dipartimento DISAT del Politecnico and Sezione INFN, Turin, Italy\\
$^{30}$ Dipartimento di Scienze MIFT, Universit\`{a} di Messina, Messina, Italy\\
$^{31}$ Dipartimento Interateneo di Fisica `M.~Merlin' and Sezione INFN, Bari, Italy\\
$^{32}$ European Organization for Nuclear Research (CERN), Geneva, Switzerland\\
$^{33}$ Faculty of Electrical Engineering, Mechanical Engineering and Naval Architecture, University of Split, Split, Croatia\\
$^{34}$ Faculty of Engineering and Science, Western Norway University of Applied Sciences, Bergen, Norway\\
$^{35}$ Faculty of Nuclear Sciences and Physical Engineering, Czech Technical University in Prague, Prague, Czech Republic\\
$^{36}$ Faculty of Physics, Sofia University, Sofia, Bulgaria\\
$^{37}$ Faculty of Science, P.J.~\v{S}af\'{a}rik University, Ko\v{s}ice, Slovak Republic\\
$^{38}$ Frankfurt Institute for Advanced Studies, Johann Wolfgang Goethe-Universit\"{a}t Frankfurt, Frankfurt, Germany\\
$^{39}$ Fudan University, Shanghai, China\\
$^{40}$ Gangneung-Wonju National University, Gangneung, Republic of Korea\\
$^{41}$ Gauhati University, Department of Physics, Guwahati, India\\
$^{42}$ Helmholtz-Institut f\"{u}r Strahlen- und Kernphysik, Rheinische Friedrich-Wilhelms-Universit\"{a}t Bonn, Bonn, Germany\\
$^{43}$ Helsinki Institute of Physics (HIP), Helsinki, Finland\\
$^{44}$ High Energy Physics Group,  Universidad Aut\'{o}noma de Puebla, Puebla, Mexico\\
$^{45}$ Horia Hulubei National Institute of Physics and Nuclear Engineering, Bucharest, Romania\\
$^{46}$ Indian Institute of Technology Bombay (IIT), Mumbai, India\\
$^{47}$ Indian Institute of Technology Indore, Indore, India\\
$^{48}$ INFN, Laboratori Nazionali di Frascati, Frascati, Italy\\
$^{49}$ INFN, Sezione di Bari, Bari, Italy\\
$^{50}$ INFN, Sezione di Bologna, Bologna, Italy\\
$^{51}$ INFN, Sezione di Cagliari, Cagliari, Italy\\
$^{52}$ INFN, Sezione di Catania, Catania, Italy\\
$^{53}$ INFN, Sezione di Padova, Padova, Italy\\
$^{54}$ INFN, Sezione di Pavia, Pavia, Italy\\
$^{55}$ INFN, Sezione di Torino, Turin, Italy\\
$^{56}$ INFN, Sezione di Trieste, Trieste, Italy\\
$^{57}$ Inha University, Incheon, Republic of Korea\\
$^{58}$ Institute for Gravitational and Subatomic Physics (GRASP), Utrecht University/Nikhef, Utrecht, Netherlands\\
$^{59}$ Institute of Experimental Physics, Slovak Academy of Sciences, Ko\v{s}ice, Slovak Republic\\
$^{60}$ Institute of Physics, Homi Bhabha National Institute, Bhubaneswar, India\\
$^{61}$ Institute of Physics of the Czech Academy of Sciences, Prague, Czech Republic\\
$^{62}$ Institute of Space Science (ISS), Bucharest, Romania\\
$^{63}$ Institut f\"{u}r Kernphysik, Johann Wolfgang Goethe-Universit\"{a}t Frankfurt, Frankfurt, Germany\\
$^{64}$ Instituto de Ciencias Nucleares, Universidad Nacional Aut\'{o}noma de M\'{e}xico, Mexico City, Mexico\\
$^{65}$ Instituto de F\'{i}sica, Universidade Federal do Rio Grande do Sul (UFRGS), Porto Alegre, Brazil\\
$^{66}$ Instituto de F\'{\i}sica, Universidad Nacional Aut\'{o}noma de M\'{e}xico, Mexico City, Mexico\\
$^{67}$ iThemba LABS, National Research Foundation, Somerset West, South Africa\\
$^{68}$ Jeonbuk National University, Jeonju, Republic of Korea\\
$^{69}$ Johann-Wolfgang-Goethe Universit\"{a}t Frankfurt Institut f\"{u}r Informatik, Fachbereich Informatik und Mathematik, Frankfurt, Germany\\
$^{70}$ Korea Institute of Science and Technology Information, Daejeon, Republic of Korea\\
$^{71}$ KTO Karatay University, Konya, Turkey\\
$^{72}$ Laboratoire de Physique des 2 Infinis, Ir\`{e}ne Joliot-Curie, Orsay, France\\
$^{73}$ Laboratoire de Physique Subatomique et de Cosmologie, Universit\'{e} Grenoble-Alpes, CNRS-IN2P3, Grenoble, France\\
$^{74}$ Lawrence Berkeley National Laboratory, Berkeley, California, United States\\
$^{75}$ Lund University Department of Physics, Division of Particle Physics, Lund, Sweden\\
$^{76}$ Nagasaki Institute of Applied Science, Nagasaki, Japan\\
$^{77}$ Nara Women{'}s University (NWU), Nara, Japan\\
$^{78}$ National and Kapodistrian University of Athens, School of Science, Department of Physics , Athens, Greece\\
$^{79}$ National Centre for Nuclear Research, Warsaw, Poland\\
$^{80}$ National Institute of Science Education and Research, Homi Bhabha National Institute, Jatni, India\\
$^{81}$ National Nuclear Research Center, Baku, Azerbaijan\\
$^{82}$ National Research and Innovation Agency - BRIN, Jakarta, Indonesia\\
$^{83}$ Niels Bohr Institute, University of Copenhagen, Copenhagen, Denmark\\
$^{84}$ Nikhef, National institute for subatomic physics, Amsterdam, Netherlands\\
$^{85}$ Nuclear Physics Group, STFC Daresbury Laboratory, Daresbury, United Kingdom\\
$^{86}$ Nuclear Physics Institute of the Czech Academy of Sciences, Husinec-\v{R}e\v{z}, Czech Republic\\
$^{87}$ Oak Ridge National Laboratory, Oak Ridge, Tennessee, United States\\
$^{88}$ Ohio State University, Columbus, Ohio, United States\\
$^{89}$ Physics department, Faculty of science, University of Zagreb, Zagreb, Croatia\\
$^{90}$ Physics Department, Panjab University, Chandigarh, India\\
$^{91}$ Physics Department, University of Jammu, Jammu, India\\
$^{92}$ Physics Department, University of Rajasthan, Jaipur, India\\
$^{93}$ Physics Program and International Institute for Sustainability with Knotted Chiral Meta Matter (SKCM2), Hiroshima University, Hiroshima, Japan\\
$^{94}$ Physikalisches Institut, Eberhard-Karls-Universit\"{a}t T\"{u}bingen, T\"{u}bingen, Germany\\
$^{95}$ Physikalisches Institut, Ruprecht-Karls-Universit\"{a}t Heidelberg, Heidelberg, Germany\\
$^{96}$ Physik Department, Technische Universit\"{a}t M\"{u}nchen, Munich, Germany\\
$^{97}$ Politecnico di Bari and Sezione INFN, Bari, Italy\\
$^{98}$ Research Division and ExtreMe Matter Institute EMMI, GSI Helmholtzzentrum f\"ur Schwerionenforschung GmbH, Darmstadt, Germany\\
$^{99}$ Saha Institute of Nuclear Physics, Homi Bhabha National Institute, Kolkata, India\\
$^{100}$ School of Physics and Astronomy, University of Birmingham, Birmingham, United Kingdom\\
$^{101}$ Secci\'{o}n F\'{\i}sica, Departamento de Ciencias, Pontificia Universidad Cat\'{o}lica del Per\'{u}, Lima, Peru\\
$^{102}$ Stefan Meyer Institut f\"{u}r Subatomare Physik (SMI), Vienna, Austria\\
$^{103}$ SUBATECH, IMT Atlantique, Nantes Universit\'{e}, CNRS-IN2P3, Nantes, France\\
$^{104}$ Suranaree University of Technology, Nakhon Ratchasima, Thailand\\
$^{105}$ Technical University of Ko\v{s}ice, Ko\v{s}ice, Slovak Republic\\
$^{106}$ The Henryk Niewodniczanski Institute of Nuclear Physics, Polish Academy of Sciences, Cracow, Poland\\
$^{107}$ The University of Texas at Austin, Austin, Texas, United States\\
$^{108}$ Universidad Aut\'{o}noma de Sinaloa, Culiac\'{a}n, Mexico\\
$^{109}$ Universidade de S\~{a}o Paulo (USP), S\~{a}o Paulo, Brazil\\
$^{110}$ Universidade Estadual de Campinas (UNICAMP), Campinas, Brazil\\
$^{111}$ Universidade Federal do ABC, Santo Andre, Brazil\\
$^{112}$ University of Cape Town, Cape Town, South Africa\\
$^{113}$ University of Houston, Houston, Texas, United States\\
$^{114}$ University of Jyv\"{a}skyl\"{a}, Jyv\"{a}skyl\"{a}, Finland\\
$^{115}$ University of Kansas, Lawrence, Kansas, United States\\
$^{116}$ University of Liverpool, Liverpool, United Kingdom\\
$^{117}$ University of Science and Technology of China, Hefei, China\\
$^{118}$ University of South-Eastern Norway, Kongsberg, Norway\\
$^{119}$ University of Tennessee, Knoxville, Tennessee, United States\\
$^{120}$ University of the Witwatersrand, Johannesburg, South Africa\\
$^{121}$ University of Tokyo, Tokyo, Japan\\
$^{122}$ University of Tsukuba, Tsukuba, Japan\\
$^{123}$ University Politehnica of Bucharest, Bucharest, Romania\\
$^{124}$ Universit\'{e} Clermont Auvergne, CNRS/IN2P3, LPC, Clermont-Ferrand, France\\
$^{125}$ Universit\'{e} de Lyon, CNRS/IN2P3, Institut de Physique des 2 Infinis de Lyon, Lyon, France\\
$^{126}$ Universit\'{e} de Strasbourg, CNRS, IPHC UMR 7178, F-67000 Strasbourg, France, Strasbourg, France\\
$^{127}$ Universit\'{e} Paris-Saclay Centre d'Etudes de Saclay (CEA), IRFU, D\'{e}partment de Physique Nucl\'{e}aire (DPhN), Saclay, France\\
$^{128}$ Universit\`{a} degli Studi di Foggia, Foggia, Italy\\
$^{129}$ Universit\`{a} del Piemonte Orientale, Vercelli, Italy\\
$^{130}$ Universit\`{a} di Brescia, Brescia, Italy\\
$^{131}$ Variable Energy Cyclotron Centre, Homi Bhabha National Institute, Kolkata, India\\
$^{132}$ Warsaw University of Technology, Warsaw, Poland\\
$^{133}$ Wayne State University, Detroit, Michigan, United States\\
$^{134}$ Westf\"{a}lische Wilhelms-Universit\"{a}t M\"{u}nster, Institut f\"{u}r Kernphysik, M\"{u}nster, Germany\\
$^{135}$ Wigner Research Centre for Physics, Budapest, Hungary\\
$^{136}$ Yale University, New Haven, Connecticut, United States\\
$^{137}$ Yonsei University, Seoul, Republic of Korea\\
$^{138}$  Zentrum  f\"{u}r Technologie und Transfer (ZTT), Worms, Germany\\
$^{139}$ Affiliated with an institute covered by a cooperation agreement with CERN\\
$^{140}$ Affiliated with an international laboratory covered by a cooperation agreement with CERN\\

\end{flushleft} 
  
\end{document}